\documentclass[
	reprint,
	article,
	twocolumn,
	nofootinbib,
	nobibnotes,
	floatfix,
	superscriptaddress
]{revtex4-2}

\usepackage{amssymb}
\usepackage{amsmath}
\usepackage{graphicx}
\usepackage{dcolumn}
\usepackage{bm}
\usepackage{hyperref}
\usepackage[usenames,dvipsnames]{xcolor}
\usepackage{listings}
\usepackage[utf8]{inputenc}
\usepackage[english]{babel}
\usepackage{booktabs}
\usepackage{hhline}

\begin{document}

\title{Interfacial Rheology of Lanthanide Binding Peptide Surfactants at the Air-Water Interface}

\author{Stephen A. Crane}
\affiliation{ 
	Department of Chemical and Biomolecular Engineering, University of Pennsylvania, Philadelphia, PA 19104, USA.
}
\author{Felipe Jimenez-Angeles}
\affiliation{ 
	Department of Materials Science and Engineering, Northwestern University, Evanston, IL 60208, USA.
}
\author{Yiming Wang}
\affiliation{ 
	Department of Chemical and Biomolecular Engineering, University of Pennsylvania, Philadelphia, PA 19104, USA.
}
\author{Luis E. Ortuno Macias}
\affiliation{ 
	Department of Chemical Engineering, The City College of New York,  New York, NY 10031, USA.
}
\affiliation{
    Levich Institute, The City College of New York,  New York, NY 10031, USA.
}
\author{Jason G. Marmorstein}
\affiliation{
    Department of Chemistry, University of Pennsylvania,  Philadelphia, PA 19104, USA.
}
\author{Jiayi Deng}
\affiliation{ 
	Department of Chemical and Biomolecular Engineering, University of Pennsylvania, Philadelphia, PA 19104, USA.
}
\author{Mehdi Molaei}
\affiliation{ 
	Department of Chemical and Biomolecular Engineering, University of Pennsylvania, Philadelphia, PA 19104, USA.
}
\author{E. James Petersson}
\affiliation{
    Department of Chemistry, University of Pennsylvania,  Philadelphia, PA 19104, USA.
}
\author{Ravi Radhakrishnan}
\affiliation{ 
	Department of Chemical and Biomolecular Engineering, University of Pennsylvania, Philadelphia, PA 19104, USA.
}
\affiliation{
    Department of Bioengineering, University of Pennsylvania,  Philadelphia, PA 19104, USA.
}
\author{Cesar de la Fuente-Nunez}
\affiliation{ 
	Department of Chemical and Biomolecular Engineering, University of Pennsylvania, Philadelphia, PA 19104, USA.
}
\affiliation{
    Department of Bioengineering, University of Pennsylvania,  Philadelphia, PA 19104, USA.
}
\affiliation{
    Machine Biology Group, Departments of Psychiatry and Microbiology, Institute for Biomedical Informatics, Institute for Translational Medicine and Therapeutics, Perelman School of Medicine, University of Pennsylvania, Philadelphia, PA 19104, USA.
}
\affiliation{
    Penn Insitute for Computational Science, University of Pennsylvania, Philadelphia, PA 19104, USA.
}
\author{ Monica Olvera de la Cruz}
\affiliation{ 
	Department of Materials Science and Engineering, Northwestern University, Evanston, IL 60208, USA.
}
\author{Raymond S. Tu}
\affiliation{ 
	Department of Chemical Engineering, The City College of New York,  New York, NY 10031, USA.
}
\author{Charles Maldarelli}
\affiliation{ 
	Department of Chemical Engineering, The City College of New York,  New York, NY 10031, USA.
}
\affiliation{
    Levich Institute, The City College of New York,  New York, NY 10031, USA.
}
\author{Ivan J. Dmochowski}
\affiliation{
    Department of Chemistry, University of Pennsylvania,  Philadelphia, PA 19104, USA.
}
\author{Kathleen J. Stebe}
\email{kstebe@seas.upenn.edu}
\affiliation{ 
	Department of Chemical and Biomolecular Engineering, University of Pennsylvania, Philadelphia, PA 19104, USA.
}

\date{April 28, 2024}

\begin{abstract}
Peptide surfactants (PEPS) are studied to capture and retain rare earth elements (REEs) at air-water interfaces to enable REE separations. Peptide sequences, designed to selectively bind REEs, depend crucially on the position of ligands within their binding loop domain. These ligands form a coordination sphere that wraps and retains the cation. We study variants of lanthanide binding tags (LBTs) designed to complex strongly with Tb$^{3+}$. The peptide LBT$^{5-}$ (with net charge -5) is known to bind Tb$^{3+}$ and adsorb with more REE cations than peptide molecules, suggesting that undesired non-specific Coulombic interactions occur. Rheological characterization of interfaces of LBT$^{5-}$ and Tb$^{3+}$ solutions reveal the formation of an interfacial gel. To probe whether this gelation reflects chelation among intact adsorbed LBT$^{5-}$:Tb$^{3+}$ complexes or destruction of the binding loop, we study a variant, LBT$^{3-}$, designed to form net neutral LBT$^{3-}$:Tb$^{3+}$ complexes. Solutions of LBT$^{3-}$ and Tb$^{3+}$ form purely viscous layers in the presence of excess Tb$^{3+}$, indicating that each peptide binds a single REE in an intact coordination sphere. We introduce the variant RR-LBT$^{3-}$ with net charge -3 and anionic ligands outside of the coordination sphere. We find that such exposed ligands promote interfacial gelation. Thus, a nuanced requirement for interfacial selectivity of PEPS is proposed: that anionic ligands outside of the coordination sphere must be avoided to prevent the non-selective recruitment of REE cations. This view is supported by simulation, including interfacial molecular dynamics simulations, and interfacial metadynamics simulations of the free energy landscape of the binding loop conformational space.
\end{abstract}

\maketitle

\section{Introduction}
We are developing peptide surfactants (PEPS) that bind rare earth elements (REEs) and adsorb at fluid interfaces for exploitation in REE recovery and purification using foam fractionation methods. The REEs, which comprise the lanthanides (Lns), Yttrium (Y), and Scandium (Sc), are considered critical elements. These elements are commonly categorized as light REEs (which include Sc, and the elements La through Gd) and heavy REEs (which include Y, and the elements Tb through Lu). Despite their name, REEs are not rare, but have crustal abundance similar to that of copper and lead, with light REEs being more abundant than heavy ones. REEs have unique properties crucial to many modern technologies including lasers, catalysts, turbines and electric vehicles.\cite{vangosen2017,long2011,goonan2011}

REE separation is challenging, particularly among the Ln cations, which are present predominantly in a +3-oxidation state and have ionic radii that differ by approximately 0.2 Å between La$^{3+}$ (the lightest Ln$^{3+}$) and Lu$^{3+}$ (the heaviest Ln$^{3+}$). This modest decrease in cation radius, the so-called lanthanide contraction, causes differences in the lanthanides’ coordination with surrounding water molecules. This contraction is also exploited to generate selectivity among extractants used in liquid-liquid extraction (LLE), the most common method for Ln separation and purification. In LLE, extractant molecules (e.g. phosphoric acids, carboxylic acids, or amines) are placed in an organic phase (typically kerosene) which is layered atop an aqueous phase that contains mixtures of REE cations. The extractants form complexes with Ln$^{3+}$ at the aqueous-organic phase interface and subsequently partition into the organic phase. The complexation of the extractants and Ln$^{3+}$ is weakly selective, allowing REEs to be separated and purified from REE mixtures.\cite{opare2021,liu2021,cheisson2019} However, neighboring REEs typically have separation factors close to 1, so many stages of LLE are required, which presents an environmental burden on regions engaged in large-scale REE separations.\cite{cheisson2019} The difficulty of these separations has inspired research on new, green alternative approaches.

\begin{figure*}
 \centering
 \includegraphics[height=7.5cm]{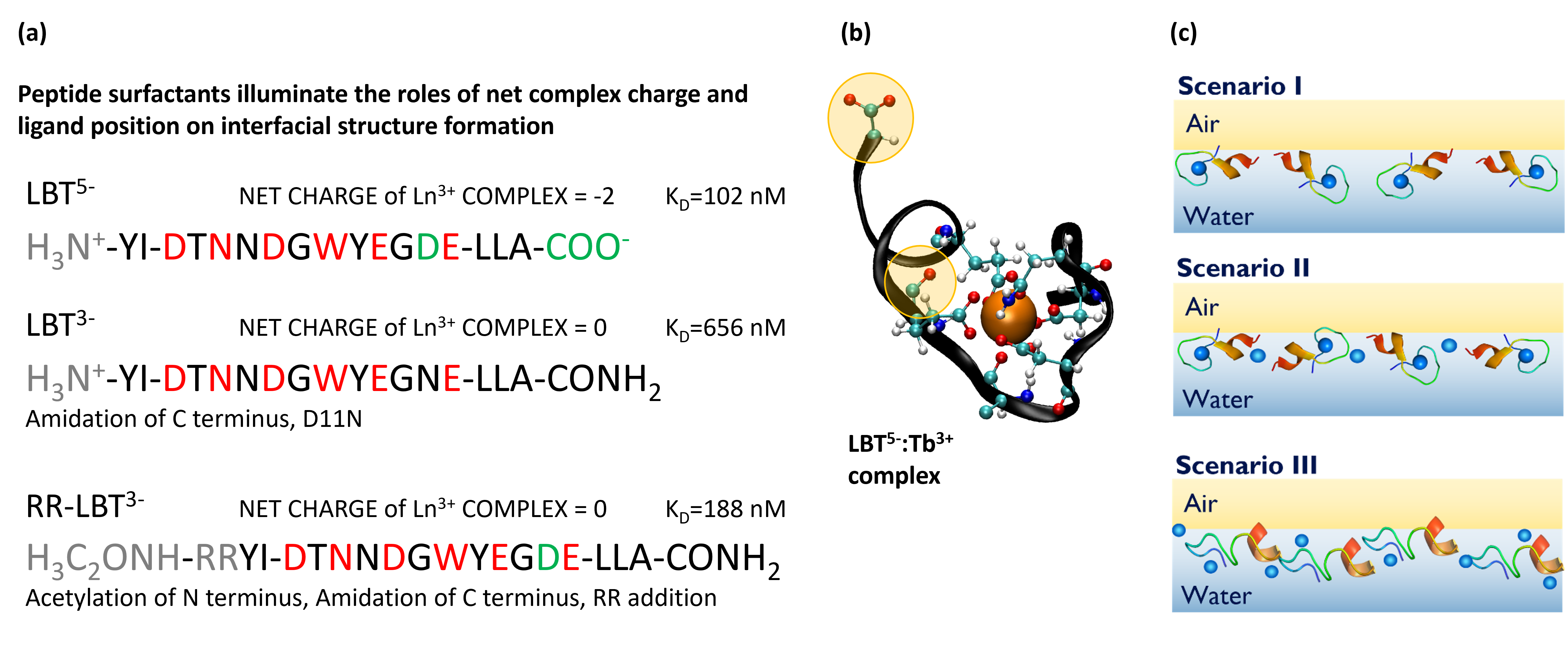}
 \caption{\textbf{PEPS sequences, structure, and hypothesized self-assembly at the air-water interface.} (a) PEPS sequences with positively charged components in gray, binding loop residues in red, and non-binding loop ligands in green. The net charge upon complexation with Tb$^{3+}$ is given along with the $K_D$ in nM for PEPS:Tb$^{3+}$ complexes. (b) An MD snapshot of the binding loop with all binding loop ligands and non-binding loop ligands shown. (c) Hypothesized PEPS morphologies at the air-aqueous interface. Depiction of PEPS acting as selective monomers with binding loop intact (Scenario I), PEPS forming multimers with intact binding loops but chelation from ligands outside the coordination sphere which reduces selectivity (Scenario II), and PEPS that denature in the air-aqueous interface resulting in complete loss of the selective binding and many PEPS:Tb$^{3+}$:PEPS interactions (Scenario III). }
 \label{fgr1}
\end{figure*}

In an interesting recent advance, peptides derived from EF-hand domains of calcium binding proteins have been immobilized on solid supports and used to capture REEs;\cite{xu2019,su2021,hostert2023} EF hands  are evolutionarily conserved, multi-ligand domains that are known to envelop a single ion in a binding loop.
 An important advantage of this approach is that REE-binding peptides based on EF-hands are relatively small biomolecules that are easy to modify and synthesize for high throughput screening.\cite{martin2005} Our work is complementary to this approach. In our envisioned process, rather than immobilizing REE-binding peptides, we aim to form selective PEPS-REE complexes in aqueous solution and to collect them at air-water interfaces in a foam recovery process. This approach exploits the enhanced and tunable selectivity of REE binding loop extractants and eliminates the organic phase of LLE which would lead to a more efficient and green process.

We develop PEPS based on lanthanide binding tags (LBTs), peptides containing a binding loop sequence of amino acids known to form a 3-dimensional coordination sphere of several (6-8) ligands, primarily carboxylates and carbonyls, around the bound REE. The LBTs have been chemically evolved from calcium binding EF hands within calcium binding proteins.\cite{martin2005,franz2003,nitz2004} Because of the similarity in size of Ca$^{2+}$ and the REE cations, these loops coordinate to REEs with even higher affinity, attributable, in part to their higher charge density. LBTs have been optimized by screening methods to coordinate with high affinity to particular lanthanides (e.g. terbium and europium). 

We have previously studied PEPS with sequences LBT$^{5-}$ and LBT$^{3-}$, shown in Figure \ref{fgr1}a; the net charge on these sequences in the unbound state is given in their superscript.  LBT$^{3-}$ is a variant of LBT$^{5-}$ in which the charge of the native peptide is altered by replacing anionic ligands outside of the coordination sphere, specifically by amidating the C terminus and exchanging an aspartic acid in the 11th position with asparagine (D11N). Figure \ref{fgr1}b shows a molecular dynamics (MD) simulation snapshot of LBT$^{5-}$ which has formed a binding loop; its backbone envelops the Ln$^{3+}$ and has its coordinating amino acid side chains facing inwardly, cradling the ion. Careful inspection of this figure shows that LBT$^{5-}$ also has two free ligands which face outside of the coordination sphere (highlighted in yellow). We refer to these ligands as non-binding loop ligands, or ligands outside the coordination sphere, which likely play a role in chelating excess ions.  To assess the importance of these ligands, we also probe the surface rheological response of the PEPS with sequence RR-LBT$^{3-}$ which has two positive arginine residues inserted before the tyrosine at the N-terminus of the LBT$^{5-}$ sequence. 

Relevant to the current study, for both LBT$^{5-}$ and LBT$^{3-}$, in prior work, we have quantified the binding affinity in bulk, and characterized air-aqueous interfaces of PEPS and Tb$^{3+}$ solutions using pendant drop tensiometry, x-ray fluorescence near total reflection (XFNTR)  and X-ray reflectivity. These data show that PEPS and Tb$^{3+}$ bind in the bulk and adsorb to form a monolayer at the interface. However, for LBT$^{5-}$ in the presence of Tb$^{3+}$ at greater than equimolar concentrations, the ratio of Tb$^{3+}$ cations to peptide present in the interface is 1.66, whereas for LBT$^{3-}$ the ratio of Tb$^{3+}$ cations to adsorbed peptide is 1.0.\cite{luis2024}  These prior results show that excess REE cation adsorption can be mitigated by managing electrostatic charge in the PEPS-REE complexes and suggest that the binding loop is intact in the interface. 

In this study, we use interfacial rheology to further probe the integrity of the LBT binding loop and, by probing judiciously selected peptide variants, show that electrostatic interactions and associated restructuring of the interfacial layer depend subtly on the location and role of ligands in the peptide sequence.  We study surface rheology at fixed PEPS concentration in the presence of Tb$^{3+}$ at concentrations varying from equimolar solutions to superequimolar solutions, and in the absence of Tb$^{3+}$. Our aim is to understand how PEPS structures relate to the ability to bind REE cations and recruit them to fluid interfaces without non-selective interactions or loss of the binding loop structure. One can imagine three scenarios for PEPS at air-aqueous interfaces, as shown schematically in Figure \ref{fgr1}c. In Scenario I, PEPS adsorb with their binding loop intact, which minimizes non-specific binding. The rheology of such interfaces is expected to be purely viscous because interactions between adsorbed PEPS:REE complexes should arise from relatively weak hydrophobic interactions. Scenario II shows PEPS that adsorb with their binding loops intact but with additional, non-selectively bound Tb$^{3+}$ ions chelated by PEPS non-coordinating ligands to form a multimeric structure. This introduces non-selective binding outside of the binding loop that would compromise the selectivity of the separation. We hypothesize that the non-binding loop ligands that face outwards play a role in this multimeric structure formation. The rheology of surfaces laden with PEPS in Scenario II would show a viscous to viscoelastic transition in the presence of excess Ln$^{3+}$ cations. Lastly, Scenario III demonstrates PEPS that denature in the interfacial layer. These PEPS would lose all selectivity and their rheology would be highly elastic, as the binding ligands of adsorbed, disordered PEPS would be accessible for crosslinking via multi-site coordination with Ln$^{3+}$. 

Interfacial shear rheology is characterized via single particle microrheology and correlated displacement velocimetry (CDV). Experimental results are corroborated by MD simulations of PEPS structural dynamics at the interface. These results allow us to infer that the binding loop is intact, albeit perturbed, in the fluid interface. Further, by probing system response in the presence of non-REE cations (e.g. Ca$^{2+}$ and Al$^{3+}$), and by study of PEPS variants designed to yield neutral PEPS-REE complexes, we propose a nuanced view of the roles of excess charge and binding ligands on the selectivity of PEPS in the interface. Our results suggest that anionic ligands located outside of the binding pocket participate in non-selective binding, whereas those within the binding loop do not.

\section{Materials and Methods}
\subsection{Sample Preparation}
Stock solutions of lyophilized peptides purchased from Genscript (>95\% purity) were prepared in MES buffer solution. Buffer was prepared by dissolving 50 mM MES (2-(N-morpholino)ethanesulfonic acid) (Acros Organics 99\%), and 100 mM NaCl, (Fisher Chemicals >99\%), in deionized water from a Milli-Q system with resistivity of 18.2 M$\Omega$ cm. The peptide concentration of each stock solution was quantified by UV-Vis spectroscopy  by measuring absorbance at 280 nm and using Beer's law with the extinction coefficient $\epsilon$=8480 cm$^{-1}$M$^{-1}$.\cite{pace1995} Solutions of Tb$^{3+}$ cations were prepared at concentrations of 5 mM from TbCl$_3\cdot$6H$_2$O (Sigma Aldrich 99.9\%) dissolved in Milli-Q water that was slightly acidified with HCl. The same procedure was followed to prepare 5 mM stock solutions of CaCl$_2$ (Fisher Chemicals $\geq$96\%), and AlCl$_3$ (Sigma Aldrich 99.99\%), respectively. Polystyrene particles suspended in solution with no surface functionalization and average diameter of 2a=1.0 $\mu$m were purchased from Polysciences. The particles were washed and centrifuged before undergoing solvent exchange with ethanol. The particles in ethanol suspensions were dried into a powder following a protocol from Bangs Laboratories, Inc. (Tech Note 203A). To ensure that particle type did not impact the results, additional control experiments were performed with both negatively charged, carboxylated particles, and positively charged, aminated particles. Experiments conducted with these particles produce similar results to the neutral particles that were used to gather the data presented in this work.

\subsection{Particle Tracking}
Particle tracking experiments were conducted in a custom-made vessel constructed by affixing a 14 mm ID ring, with a lower half made of aluminum and upper half made of Teflon, to a No. 1 coverslip using Corning vacuum grease to allow for imaging on an inverted microscope.\cite{lee2010,deng2020} Care is taken to ensure a good seal without introducing grease to the sample. Prior to affixing to the ring, the coverslip is treated with O$_2$ plasma, making the glass surface more hydrophilic, which aids in forming a flat air-aqueous interface. Appropriate dilutions of Tb$^{3+}$ stock solutions are made in buffer and placed in the vessel, in which they spread to form a planar interface. Colloidal particles are introduced to the interface via an aerosolization method to avoid the use of spreading solvents. In this method, clusters of dried particles are rubbed between two clean microscope slides to form a thin layer of individual particles. The dried particles are aerosolized using a blast of compressed gas and allowed to settle under gravity onto the interface.\cite{gharbi2011} The peptide solution is gently added by injection beneath the interface after the particles to fill the sample chamber to the aluminum-Teflon seam, with a total sample volume of 120 $\mu$L. The Teflon seam pins the air-liquid interface, which has an area of  A$_{surface}$=1.54 cm$^2$. Unless otherwise noted, all particle-tracking experiments were conducted on solutions of 65 µM PEPS with varying molar concentrations of Tb$^{3+}$ ranging from 0 $\mu$M to 260 $\mu$M, i.e. spanning a molar ratio of Tb$^{3+}$ to PEPS from 0.0 to 4.0. Samples are allowed to equilibrate for 30 min before particle tracking begins. Particle tracking is carried out using an inverted microscope in brightfield with a 40x air objective (NA 0.55). For each condition studied, three videos of 5000 frames captured at 26 frames per second are recorded at three distinct positions at the interface using a CMOS camera (Point Grey). The images are analyzed using a Python implementation of the algorithm by Crocker et al. to track each particle’s position.\cite{crocker1996,allan_2024_10696534} Collective drift of all particles is removed.

Two methods are used to characterize surface shear rheology. For purely viscous interfaces, we determine the surface viscosity $\eta_s$, which we report in non-dimensional form in terms of the Boussinesq number,
\begin{equation}
  Bq = \frac{\eta_s}{\eta a}
  \label{eq1}
\end{equation}
which characterizes the importance of surface viscous effects relative to viscous effects in the bulk phase. In this expression, $\eta$ is the shear viscosity of the solution, and $a$ is the particle radius. For all experiments in this study, we have used surface probes with radius $a$=0.5 $\mu$m. We take the value of the bulk fluid viscosity to be that of water, $\eta$=0.001 Pa$\cdot$s, so Bq=1.0 for surface viscosity $\eta_s$=0.5 nP$\cdot$m$\cdot$s.  

For viscoelastic interfaces, we determine the interfacial storage and loss moduli, G$^{\prime}$ and G$^{\prime\prime}$, respectively.  We report the storage modulus in terms of $\hat{G}$, a non-dimensional number defined in terms as a ratio of characteristic interfacial elastic to bulk viscous effects,
\begin{equation}
  \hat{G} = \frac{G'}{\eta a \omega|_{\omega=1}}
  \label{eq2}
\end{equation}
where the characteristic bulk viscous scaling is defined at frequency  $\omega$=1.0 s$^{-1}$.

\subsubsection{Single point interfacial microrheology.}
To perform single point microrheology, the mean squared displacement (MSD) of each particle is calculated according to
\begin{equation}
  \langle\Delta r^2(\tau)\rangle = \langle[\mathbf{r}(t+\tau)-\mathbf{r}(t)]^2\rangle
  \label{eq3}
\end{equation}
where $\tau$ is the lag time, and $\mathbf{r}$ is the position of the particle in the interface at a particular time. The resolution of particle tracking is limited by noise and vibrations. To determine the resolution of particle displacements in our laboratory, the MSD of the perceived “motion” of static particles deposited on a glass coverslip is tracked. This value, approximately 0.003 $\mu$m$^2$, provides a lower bound to the MSD measurements (Fig. S1).

To infer local surface rheology or to probe the heterogeneity of response on the interface, the MSD of individual colloidal particles can be tracked. Alternatively, to infer average surface rheological response of the interface, the ensemble averaged mean square displacements (EMSD) can be analyzed. In either case, the MSD (or EMSD) vs. $\tau$ is fitted with a power-law model,
\begin{equation}
  \langle\Delta r^2(\tau)\rangle = A \tau ^n
  \label{eq4}
\end{equation}
where $A$=4$D$, with $D$ being the diffusivity of the particle and $n$=1 for an interface that behaves as a two-dimensional viscous fluid. For such interfacial films, the particle diffusivity inferred from this fit is used to calculate the surface viscosity using the Einstein relation, which relates the surface diffusion coefficient of a colloidal probe to the mobility (terminal velocity per unit force) of a probe particle in the interface, which depends on the drag coefficient on the particle. We adopt the drag coefficient calculated for a disk in a thin viscous interfacial film that does not penetrate the surrounding fluid phases which is valid over a broad range of Bq\cite{saffman1975,saffman1976,hughes1981} which has the form:
\begin{equation}
  D = \frac{k_B T}{4\pi \eta a \Lambda(Bq)}
  \label{eq5}
\end{equation}
where $\Lambda$ is a dimensionless drag coefficient that is given by:
\begin{equation}
  \Lambda(Bq) = [\frac{1}{Bq}(\ln{2Bq}-\gamma+\frac{4}{\pi Bq}-\frac{1}{2 Bq^2}\ln{2Bq}+\mathit{O}(\frac{1}{Bq^2}))]
  \label{eq6}
\end{equation}
where $\gamma$ is the Euler-Mascheroni constant. Equations \ref{eq5} and \ref{eq6} allow Bq to be determined from a measured diffusivity. Our results are not highly dependent on this choice of drag coefficient; a plot of $D$ vs Bq as given by equations \ref{eq5} and \ref{eq6} can be found in Fig. S2. An alternative drag coefficient that accounts for the contact angle of a partially submerged sphere\cite{fischer2006} is within a factor of 3 for most contact angles and is also shown in Fig. S2. In our measurements, we find Bq$\gg$1 for many of the viscous interfacial films.  In this regime, the differences between the various drag coefficient formulations are negligible, and all agree with the classic result of Saffman\cite{saffman1975,saffman1976} for highly viscous surface films.

Interfaces with n < 1 have viscoelastic behavior and are described by a complex surface modulus. The storage and loss modulus (G$^{\prime}$ and G$^{\prime\prime}$ ) are calculated from MSD (or EMSD) vs. $\tau$ data using the equation,
\begin{equation}
  \Tilde{r}^2(s) = \frac{k_B T}{\pi s a \Tilde{G}(s)}
  \label{eq7}
\end{equation}
where  $\Tilde{r}^2(s)$ is the Laplace transform of the MSD, s is the Laplace frequency, and $\Tilde{G}(s)$ is the modulus in Laplace space.\cite{mason1995} Local power law fits, determined by the logarithmic time derivative of the MSD, are used to estimate the complex shear modulus. We use a second-order polynomial to locally smooth the MSD data and to numerically calculate the first- and second-order logarithmic derivatives of the MSD. By including the second-order logarithmic time derivatives of the MSD, we achieve a better estimate of the moduli.\cite{dasgupta2002}

Absent PEPS, single point microrheology of the air-aqueous interfaces of buffer solutions and of buffer solutions containing 260 $\mu$M Tb$^{3+}$ are viscous, with Bq $\sim$ 2 and Bq $\sim$ 5, respectively, indicating the presence of surface-active impurities in the buffer and Tb$^{3+}$ components (Fig. S3).

\subsubsection{Correlated displacement velocimetry.}
Brownian displacement of colloidal particles occurs with negligible inertia; in this limit, the displacement of any particle instantaneously generates a flow field that displaces neighboring particles. The spatial form of this flow field reveals the stress state of the fluid interface. The flow field is imaged using correlated displacement velocimetry (CDV)\cite{molaei2021} in which  the correlated displacement of particle pairs at various relative positions on the interface is calculated and superposed to reveal the flow. The method is described in detail in Molaei et al.\cite{molaei2021}; the main concepts are presented here. In CDV, a particle on the interface is selected as a displacement source (denoted “s”) and neighboring particles are considered as probe particles (denoted “p”). Centered on the source particle, a Cartesian coordinate system (x,y) is constructed, with y corresponding to the direction of source displacement over lag time $\tau$. The position of the probe particle with respect to the source is given by $\mathbf{x}_{sp}$. The correlated displacement vector for the source-probe particle pair is calculated over lag time $\tau$, given by the product of the source particle’s displacement $\Delta y(t,\tau)$ and the displacement of the probe in the interface $\Delta \mathbf{x}^p(t,\tau)$. This process is repeated, with each particle on the interface playing the role of source particle. The correlated displacement vector field $\mathbf{\chi}_{ij}$ is calculated for the i$^{th}$ source particle’s displacement $\Delta y_i(t,\tau)$ and the j$^{th}$ probe particle’s displacement, $\Delta \mathbf{x}_j^p(t,\tau)$, and is averaged over all particle-tracer pairs for $\mathbf{x}_{sp}$ equal to $\mathbf{x}$:
\begin{multline}
  \mathbf{\chi}(\mathbf{x},\tau)=\langle\mathbf{\chi}_{ij}(\mathbf{x}_{sp}(t,\tau))\rangle=\\ \langle\Delta y_i(t,\tau)\Delta \mathbf{x}_j^p(t,\tau)\delta^{2D}(\mathbf{x}-\mathbf{x}_{sp}(t))\rangle_{t,i,j}
  \label{eq8}
\end{multline}
In this expression, the brackets $\langle \ldots \rangle_{t,i,j}$ indicate averaging over time and particle pairs and $\delta^{2D}(\mathbf{x}-\mathbf{x}_{sp}(t))$ is a 2D Dirac delta function. Given that, over lag time $\tau$, the ensemble averaged displacement of the source colloids is $\langle\Delta r^2(\tau)\rangle^{1/2}$, and a probe particle in the flow field $\mathbf{u}(\mathbf{x})$ moves a distance $\mathbf{u}(\mathbf{x})\tau$, Eq. \ref{eq8} can be related to the velocity field:
\begin{equation}
  \mathbf{u}(\mathbf{x}) = \frac{\mathbf{\chi}(\mathbf{x},\tau)}{\langle\Delta r^2(\tau)\rangle^{1/2}\tau}
  \label{eq9}
\end{equation}
To measure the flow field, all frames of source-probe displacements over lag time  are appropriately superposed in the (x,y) frame.  Probe displacements are binned so signal can be discerned relative to noise.  We experiment with bin sizes so that each bin has a sufficient number of data points to resolve signal over the noise (Fig. S4). We approximate the correlation vector for $\mathbf{x}$ in the k$^{th}$ bin as $\mathbf{\chi}^k = \langle\Delta y_i(\tau)\Delta\mathbf{x}^p_{j,k}(\tau)\rangle_{i,j}^k $ where the averaging is performed over all probes located in the region R$^k$. The mean velocity disturbance in each bin is then found as $\mathbf{u}(\mathbf{x}\in R^k) \approx \frac{\mathbf{\chi}^k}{\langle\Delta r^2(\tau)\rangle^{1/2}\tau}$. For purely elastic interfaces, the same approach yields the shear strain field generated by the probe’s Brownian displacement in the film $\mathbf{U}(\mathbf{x}\in R^k) \approx \frac{\mathbf{\chi}^k}{\langle\Delta r^2(\tau)\rangle^{1/2}}$.

For fluid interfaces, the spatial form of the flow field reveals an interfacial Stokeslet generated by “Brownian forcing” by a point force $\mathit{\mathbf{f}}$ directed along the y axis in an incompressible fluid interface. Surface viscosities slow the rate of spatial decay of the flow and generate signatures that can be used to quantify the surface viscosity. In particular, the y-directed velocities along the x and y axis depend strongly on the surface viscosity and can be fitted with the theoretical flow field using two parameters: the point force, $\mathit{\mathbf{f}}$, and the Boussinesq length $L_B$.  The theoretical y component of the velocity u$_y$ for a point force acting in the interface is given by:
\begin{equation}
  u_y=\mathbf{f}(\frac{\Phi_0-\Phi_2}{4\pi\eta}+\frac{\Phi_2 y^2}{2\pi\eta|\mathbf{x}|^2})
  \label{eq10}
\end{equation}
with
\begin{equation}
  \Phi_n = \int_0^\infty \frac{1}{2+L_Bk} J_n(ks)dk
  \label{eq11}
\end{equation}
where the Boussinesq length $L_B=\eta_s/\eta$ and $J_n$ is the Bessel function of the first kind of order n.\cite{chisholm2021}
Careful drift removal and symmetrization is performed as described in the SI (Fig. S5). The y-directed velocities on the y and x axes are also related to the two-point correlation functions $D_{rr}$ and $D_{\theta\theta}$, respectively. The function $D_{rr}$ measures correlated motion along the line joining the centers of particles, and $D_{\theta\theta}$ measures correlated motion perpendicular to the line joining the centers of particles, like the y-directed velocity along the x-axis. These correlation functions are divided by lag-time $\tau$ to characterize surface rheology in the method known as two-point microrheology.\cite{stone1998,prasad2006} Inspection of the form of the CDV velocity shows that we can recover the two-point correlation functions by multiplying $u_y\langle\Delta r^2(\tau)\rangle^{1/2}$. Using CDV, we recover expected asymptotic behavior for $L_B\gg1$, i.e. 
$u_y|_{(x=0)}$  decays as $1/|\mathbf{x}|^2$, and $u_y|_{(y=0)}$ decays as $1/|\mathbf{x}|$ (Fig. S6).

\subsection{Interfacial Tensiometry}
The surface tension of PEPS:REE solutions is measured using pendant drop tensiometry in which a  silhouette of a pendant droplet of solution is fitted with the Young-Laplace equation.\cite{rotenberg1983} Pendant droplets are formed at the end of an 18-gauge straight needle by a syringe and kept in a humidified chamber which limits evaporative losses. Silhouettes are captured using a camera (Basler acA720-520um) every 5 seconds for a total of one hour. The surface tension of clean air-water interfaces ($\gamma$ = 72.8 mN/m) is measured each time before a PEPS and REE mixture is measured to ensure cleanliness of the syringe and needle and to confirm proper calibration of the instrument. For peptide and REE solutions that do not come to equilibrium, but have continued weak relaxation of tension, we truncate this measurement at 45 minutes and report the surface tension values at that surface age as quasi-equilibrium surface tensions. For elastic interfaces, we report the apparent surface tension obtained from the pendant drop silhouette as an apparent surface tension, noting that its interpretation must be treated with care as the interface is covered in a thin solid film.

\subsection{Molecular Dynamics}
Molecular dynamics (MD) simulations were performed for the two peptides LBT$^{5-}$ and LBT$^{3-}$. The conformation of LBT$^{5-}$ is taken from the protein data bank (ID 1TJB).\cite{nitz2004} To simulate the interfacial behavior, we employ a simulation box with an aqueous phase slab\cite{jimenez2019} at the middle of the simulation box. The aqueous slab is surrounded by empty spaces which accommodate the vapor phase and create two surfaces (Fig. S7a). The simulation box dimensions are 10 nm $\times$ 10 nm $\times$ 40 nm in the x-, y-, and z-directions, respectively. We applied periodic boundary conditions in the three directions. At t = 0, a layer of PEPS complexes is placed at each of the interfaces in the configuration shown in Fig. S7a. The systems are simulated for at least 1 $\mu$s and up to 2 $\mu$s. We investigate the conformational evolution of the peptide complexes, the velocity profiles, and mean square displacement. We considered variable complex concentrations. The composition of the systems is given in Table S1. We conducted classical all-atom molecular dynamics (MD) simulations using the package GROMACS.\cite{pall2020,hess2008,abraham2015} The electrostatic and van-der-Waals interactions are considered using the parameters from the CHARMM force field.\cite{brooks2009} The electrostatic interactions are computed using the PME algorithm. We performed microsecond-MD simulations in the NVT ensemble at T = 298 K using a time-step of 2.5 fs.  The temperature is controlled using the Nose-Hoover thermostat.\cite{nose1984,hoover1985} The systems are equilibrated for 20 ns before a production run of at least 1 $\mu$s and up to 2 $\mu$s. Two and three replicated simulations are conducted for the systems. To calculate the mean square displacement, the system configurations are recorded every 20 ps. No significant differences are found by recording every 100 ps.

\subsection{Metadynamics Simulation}
Well-tempered metadynamics simulations were performed to construct the free energy landscape of the LBT$^{5-}$:Tb$^{3+}$ and LBT$^{3-}$:Tb$^{3+}$ complex at the air-water interface using the GROMACS package 2020.2\cite{abraham2015} and Plumed 2.6.2 version.\cite{bonomi2009} The initial configuration of the LBT$^{5-}$:Tb$^{3+}$ complex was obtained from X-ray measurements (PDB code: 1TJB).\cite{nitz2004} The initial configuration of the LBT$^{3-}$:Tb$^{3+}$ complex was obtained by residue mutation D11N using the Scwrl4 program.\cite{krivov2009} Metadynamics simulations of the two systems were performed under constant volume (rectangular box with 5 nm $\times$ 5 nm $\times$ 10 nm in XYZ direction, respectively) and the room temperature maintained by the stochastic velocity rescaling thermostat.\cite{bussi2007} The initial configurations for the two metadynamics simulations are obtained from the last configuration of 100 ns NVT simulations of LBT$^{5-}$:Tb$^{3+}$ complex and LBT$^{3-}$:Tb$^{3+}$ complex at the air-water interface. Here, we used two collective variables (CVs). First, the coordination number (CN) of Tb$^{3+}$, which refers to the number of oxygen atoms (j) from the LBT peptide that have a separation distance with Tb$^{3+}$ (i) less than $r_c$ = 0.27 nm and do not include the oxygens of water, using the following equation:
\begin{equation}
  CN=\sum_j \frac{1-(r_{ij}/r_c)^{16}}{1-(r_{ij}/r_c)^{32}}
  \label{eq12}
\end{equation}
Second, the root mean square deviation of all $\alpha$ carbons (C$_{\alpha}$-RMSD) coordinate $R_i$ (carbon C$_i$, i = 1-12) of the 3$^{rd}$-14$^{th}$ residues (corresponding to position 1 to 12) with respect to the reference structure coordinate $R_{ref}$, (PDB code: 1TJB)\cite{nitz2004}, using the following equation:
\begin{equation}
  RMSD = \frac{1}{N}\sum_i\sqrt{(R_i-R_{ref})^2}
  \label{eq13}
\end{equation}
The height of the Gaussian potential was 1.0 kJ/mol which was deposited every 500 steps (PACE). The widths (SIGMA) of Gaussian potentials for CN and C$_{\alpha}$-RMSD were 0.3 and 0.05, respectively. The bias factor for well-tempered metadynamics was set to be 3. In addition, the upper wall restraining potential was imposed at C$_{\alpha}$-RMSD = 0.6 nm and the lower wall restraining potential was imposed at CN = 5. Both upper and lower wall potential used a force constant (KAPPA) of 300 kJ and a power (EXP) of 2. The grid boundaries (GRIDMIN and GRIDMAX) for CN and C$_{\alpha}$-RMSD were set to be 2 $\sim$ 11 and 0 $\sim$ 1.1 nm, respectively. The number of bins for every collective variable (GRIDBIN) was set to be 500. Eight replicas with different random seeds were run in parallel with each replicate lasting around 2 $\mu$s.

\section{Results and Discussion}
\subsection{Interfacial rheology of LBT$^{5-}$ and Tb$^{3+}$ solutions}
\begin{figure}[h]
 \centering
 \includegraphics[height=8cm]{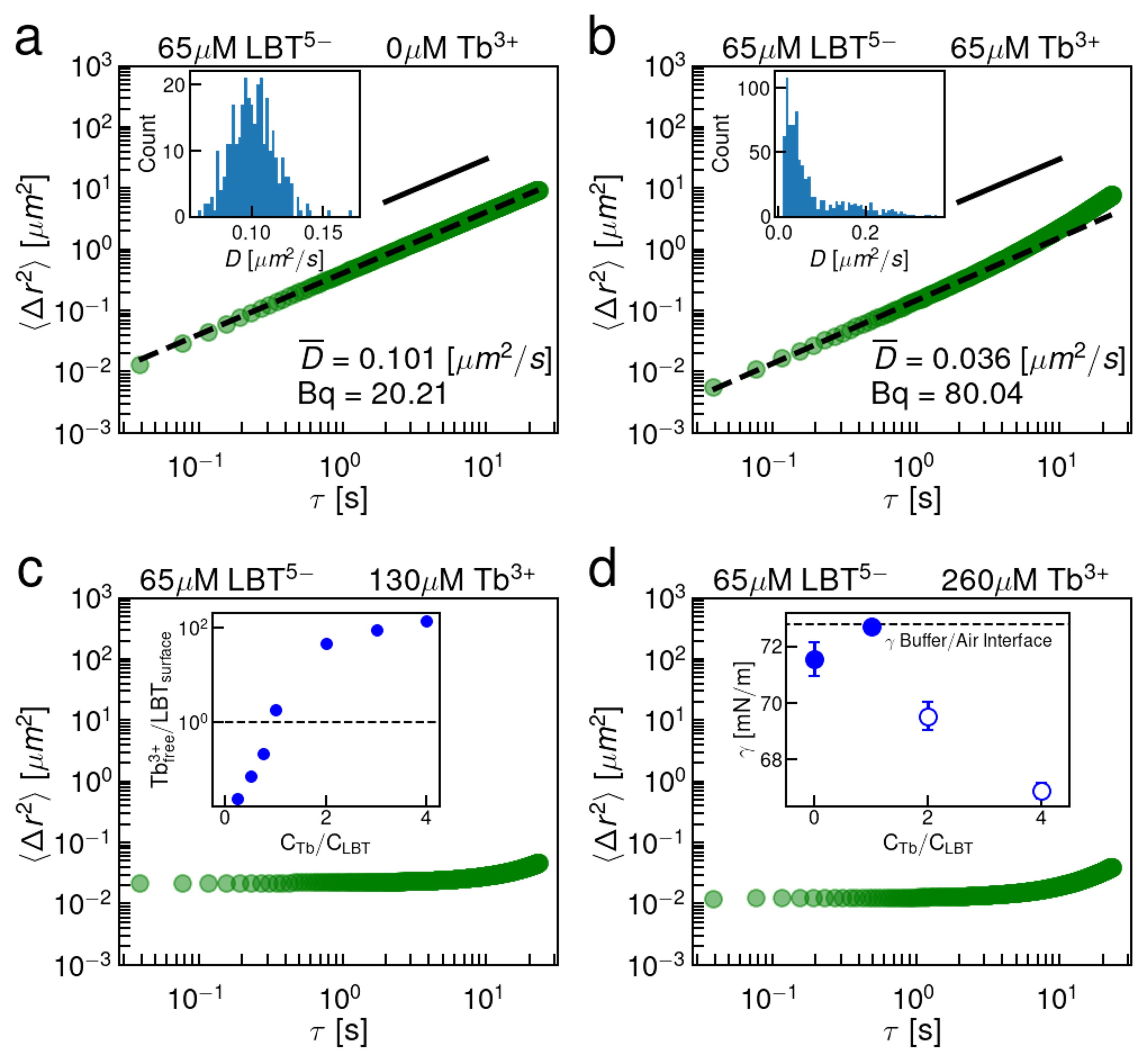}
 \caption{\textbf{Interfaces of solutions of LBT$^{5-}$ and Tb$^{3+}$ are elastic above equimolar concentrations of Tb$^{3+}$.} EMSD of probe particles (2a=1 $\mu$m) at the air-water interface in the presence of bulk solution containing (a) 65 $\mu$M LBT$^{5-}$ and 0 $\mu$M Tb$^{3+}$; (b) 65 $\mu$M LBT$^{5-}$ and 65 $\mu$M Tb$^{3+}$; (c) 65 $\mu$M LBT$^{5-}$ and 130 $\mu$M Tb$^{3+}$; (d) 65 $\mu$M LBT$^{5-}$ and 260 $\mu$M Tb$^{3+}$. Insets in (a) and (b) show the distribution of particle diffusivities. The black dashed lines are fits of the form MSD=4D$\tau$. The solid black lines are guides to show a power law with an exponent of 1.0. Inset in (c) is a calculated amount of free Tb$^{3+}$ after binding to LBT$^{5-}$ divided by the amount of LBT$^{5-}$ that could maximally pack at the interface versus the ratio of Tb$^{3+}$ to LBT$^{5-}$ in solution. Inset in (d) is the surface tension after 45 minutes for each solution. The filled symbols represent the surface tension of the viscous droplets, while the open symbols show an apparent tension in the elastic PEPS layer.}
 \label{fgr2}
\end{figure}

Interfaces of LBT$^{5-}$ solutions with excess Tb$^{3+}$ cations above an equimolar concentration ratio undergo a viscous to elastic transition, as is evident in the single point microrheology data shown in Figure \ref{fgr2}.  Absent Tb$^{3+}$, the LBT$^{5-}$ laden interface is highly viscous, with Bq $\sim$ 20 (Fig. \ref{fgr2}a); the significant dissipation in the interface may be attributed to the intrinsically disordered unbound peptide and resulting peptide-peptide interactions. For equimolar Tb$^{3+}$ to LBT$^{5-}$ concentrations, the interface is even more sluggish, with Bq $\sim$ 80 (Fig. \ref{fgr2}b). For excess Tb$^{3+}$ (Fig. \ref{fgr2}c) the EMSD shows the signature of elastic network formation. The nearly constant reduced EMSD at early lag times indicates that surface probes are caged within the network. The increase in EMSD at long lag times indicates that the probes can escape from these caging domains. As even more Tb$^{3+}$ is added (Fig. \ref{fgr2}d), the elastic network stiffens as indicated by the slight reduction in the MSD plateau. We find the storage modulus G$^{\prime}(\omega$=1 s$^{-1}$) = 82.9 mN/m and loss modulus G$^{\prime\prime}(\omega$=1 s$^{-1}$) = 15.4 mN/m; G$^{\prime}$ > G$^{\prime\prime}$, indicating a transition to a thin solid film. 

\begin{figure*}
 \centering
 \includegraphics[height=9.cm]{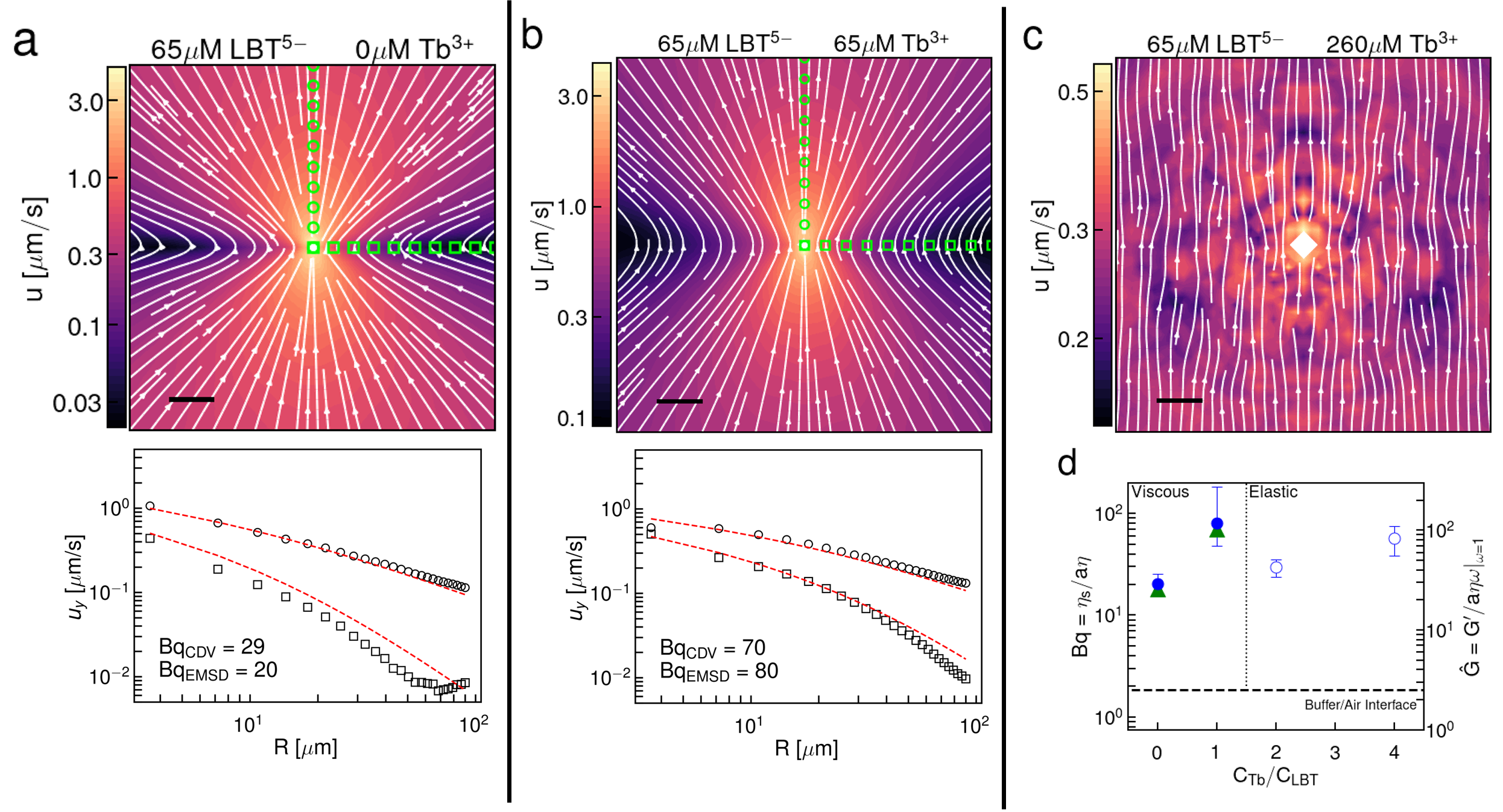}
 \caption{\textbf{Interfacial flow fields around Brownian colloids corroborate surface viscosity of single point measurements. }Flow fields generated from the Brownian motion of the probe particles at the air-aqueous interface of solutions containing (a) 65 $\mu$M LBT$^{5-}$ and 0 $\mu$M Tb$^{3+}$; (b) 65 $\mu$M LBT$^{5-}$ and 65 $\mu$M Tb$^{3+}$; (c) 65 $\mu$M LBT$^{5-}$ and 260 $\mu$M Tb$^{3+}$ (scale bar = 20 $\mu$m). Y-directed velocities along the y-axis (circles) and x-axis (squares) as measured by CDV for the viscous interfaces shown below. Theoretical velocity fits are shown by a red dashed line. (d) Summary of shear rheological behavior with single point Bq shown in filled blue symbols and Bq measured by CDV in filled green symbols (left axis). Non-dimensional storage modulus $\hat{G}$ from single point measurements is shown in open blue circles (right axis).}
 \label{fgr3}
\end{figure*}
This crossover can be related to the presence of excess unbound cations in solution that are not in binding loops.  Molecular dynamics simulation of LBT$^{5-}$:Tb$^{3+}$ complexes in the interface yields a minimum area per molecule $a_{min}=150\  \text{\AA}^2$/molecule. Using this number and the known interfacial area of our chamber $A_{surface}$, we estimate the maximum number of moles of complexed peptide that can adsorb at the interface of our chamber, $N^{surface}_{complex}=A_{surface}/a_{min}$.  Using the binding affinity of the LBT$^{5-}$:Tb$^{3+}$ complex in solution ($K_D$=150 nM), we also estimate the total unbound fraction of Tb$^{3+}$ in solution $N_{Tb^{free}}$. For fixed peptide concentration, the ratio  $N_{Tb^{free}}$/$N^{surface}_{complex}$, shown in the inset to Fig. \ref{fgr2}c, increases above unity (dashed line) for bulk concentrations of Tb$^{3+}$ in excess of equimolar ratios to peptide. We hypothesize that these excess cations interact with the adsorbed LBT$^{5-}$:Tb$^{3+}$ complexes to form condensed domains of multimers of LBT$^{5-}$:Tb$^{3+}$ complexes that are chelated by Tb$^{3+}$ through free anionic ligands present on the PEPS:REE complexes.

Using the same videos of Brownian particles in the interface, we also perform CDV and construct the interfacial flow generated around a Brownian colloid in the interface (Fig. \ref{fgr3}a-c). The flow field has the form of a Stokeslet in a viscous, incompressible interface.\cite{molaei2021,chisholm2021} In the presence of excess Tb$^{3+}$ cations, the displacement field in the thin elastic film is nearly uniform (Fig. \ref{fgr3}c), corresponding to that of a source particle generating unidirectional displacement of surface probes. These data corroborate the main single-point rheology result, i.e. that the interface transitions from a viscous fluid to a thin elastic solid film in the presence of excess Tb$^{3+}$. Moreover, the measured values of Bq from CDV closely resemble the values measured by single point microrheology (Fig. \ref{fgr3}d). This result is notable because CDV does not probe the interface in the direct vicinity of the particle, rather it reports on the region between two particles. This means the method is insensitive to any local interfacial structuring that could be brought on by the probe particle.\cite{crocker2000} The spatial form of the flow fields generated by Brownian colloids constructed by CDV for viscous interfaces (Fig. \ref{fgr3}a, b) clearly shows that there is an incompressible layer, as expected for interfaces that are laden with surface-active species that do not ad/desorb rapidly between the interface and the fluid sublayer. The waist of the hourglass shape of the flow field expands with increasing surface viscosity and the velocity decays begin at distances further from the origin. We fit the y-directed velocities along the x and y axes to extract a Bq from the viscous flow fields. Theoretical flow fields and y-directed velocity profiles along the x and y axis are shown in Fig. S6. 

In the above discussion, we have hypothesized that the elastic interface indicates non-specific binding in the interface by chelation of Tb$^{3+}$ ions that link complexed LBT$^{5-}$:Tb$^{3+}$ in the interface. In this scenario, the coordination sphere of binding ligands forms and envelops a single Tb$^{3+}$ cation, but the additional anionic ligands present outside of the coordination sphere interact with excess Tb$^{3+}$ to form viscous aggregates and elastic networks. Alternatively, the binding loop could denature in the interface, thereby presenting many anionic ligands for network formation. Such denaturation could arise from two sources: the anisotropic environment of the air-aqueous interface itself, or electrostatic interaction between net negative complexes and excess Tb$^{3+}$ ions. 

\subsection{Interfacial rheology of LBT$^{3-}$ and Tb$^{3+}$ solutions}
\begin{figure}[h]
 \centering
 \includegraphics[height=8cm]{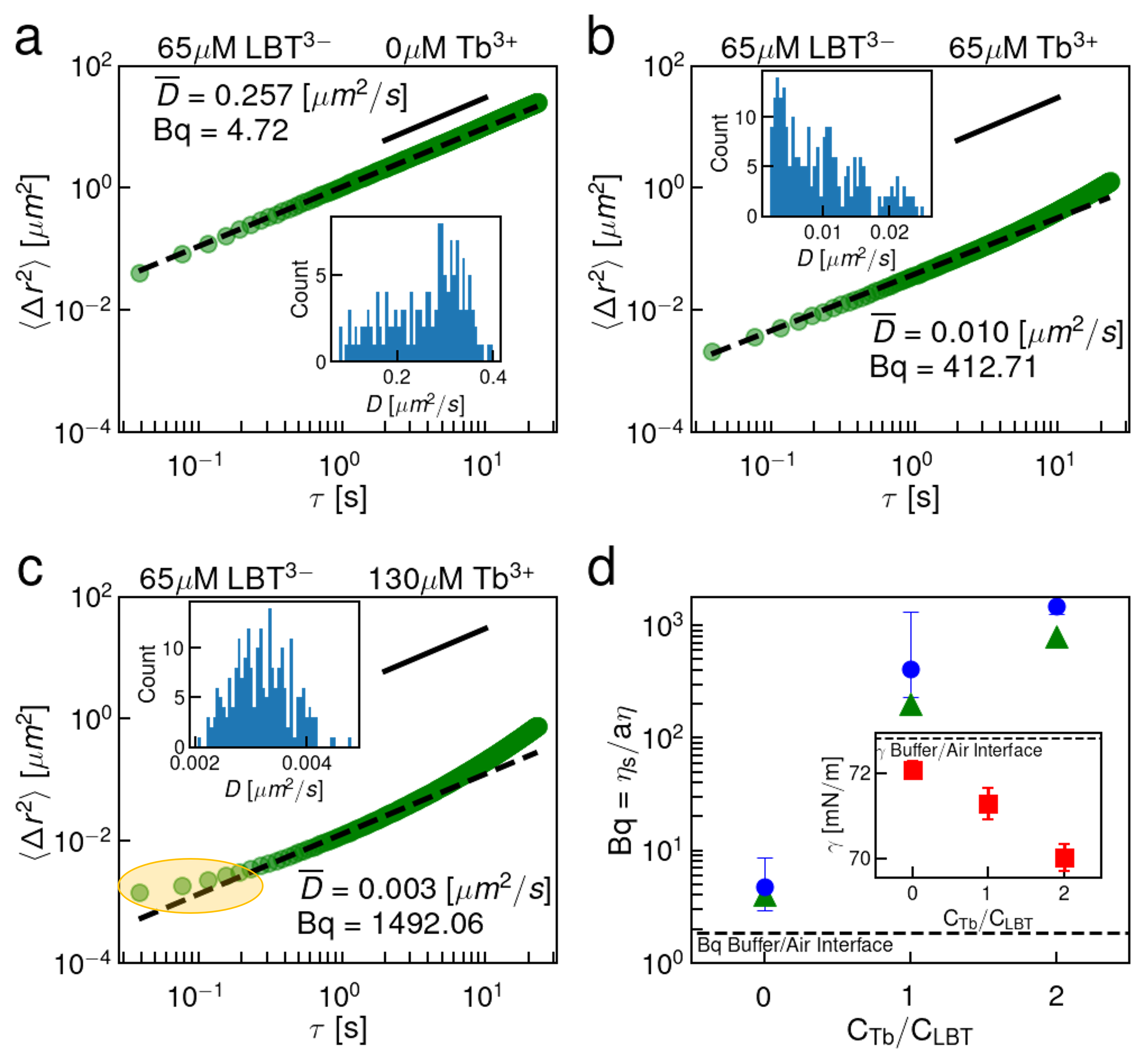}
 \caption{\textbf{Interfaces of solutions containing LBT$^{3-}$ and Tb$^{3+}$ are purely viscous.} EMSD of (2a=1.0 $\mu$m) tracer particles at the air-water interface in the presence of bulk solution containing (a) 65 $\mu$M LBT$^{3-}$ and 0 $\mu$M Tb$^{3+}$; (b) 65 $\mu$M LBT$^{3-}$ and 65 $\mu$M Tb$^{3+}$; (c) 65 $\mu$M LBT$^{3-}$ and 130 $\mu$M Tb$^{3+}$. Inset is the distribution of particle diffusivities showing a relatively unimodal distribution of diffusivities. The black dashed lines are fits of the form MSD = 4D$\tau$. The solid black lines are guides to show a power law with an exponent of 1.0. (d) A summary of Bq vs lag-time for single point (blue circles) and CDV (green triangles) measurements. Inset in (d) is the quasi-equilibrated surface tension.}
 \label{fgr4}
\end{figure}

To  assess the binding loop integrity, a less charged peptide surfactant, LBT$^{3-}$, was selected. This peptide has identical coordinating ligands to LBT$^{5-}$ but forms neutral complexes with Tb$^{3+}$ and contains no anionic ligands outside of its coordination sphere. Figure \ref{fgr4} summarizes the interfacial rheology characterization of LBT$^{3-}$. The EMSD data, shown in Figure \ref{fgr4}, show that the LBT$^{3-}$ and Tb$^{3+}$ laden interfaces are viscous at all concentrations of Tb$^{3+}$, over concentration ratios of Tb$^{3+}$ to LBT$^{3-}$ from 0 to 2.0. Furthermore, the distributions of tracer diffusivities measured from individual particles’ MSDs are relatively unimodal, indicating that interfacial heterogeneity is not pronounced. The viscosity of these interfaces becomes extremely large with excess Tb$^{3+}$. This highly dissipative behavior is likely driven by van der Waals interactions among the neutral LBT$^{3-}$:Tb$^{3+}$ complexes which have limited electrostatic repulsion. A careful scrutiny of the EMSD at Tb$^{3+}$ to LBT$^{3-}$ concentration ratio of 2 shows an apparent plateau in EMSD at short lag times (Fig. \ref{fgr4}c, highlighted). However, this plateau does not indicate the formation of an elastic film.  Rather, the viscosity in this interface is so high in these layers that the particle displacement cannot be measured due to the resolution of the instrument (Fig. S1). At lagtimes of 0.2 s and longer, probe motion can be resolved and the EMSD is diffusive. The surface tension of these solutions also differs from that of LBT$^{5-}$ solutions (Fig. \ref{fgr2}d inset); the surface tension decreases monotonically over all Tb$^{3+}$ to LBT$^{3-}$ concentration ratios (Fig. \ref{fgr4}d inset). While for LBT$^{5-}$, electrostatic repulsion was significantly reduced by chelation of excess cations, such effects are absent for LBT$^{3-}$ which lacks excess charge when complexed to drive Tb$^{3+}$ chelation. A summary of Bq versus Tb$^{3+}$ concentration (Fig. \ref{fgr4}d) shows that the surface viscosity increases monotonically with Tb$^{3+}$ and appears to plateau for Tb$^{3+}$ in excess of equimolar ratios. Again, CDV was performed on these interfaces (Fig. S8) and the Bq from CDV follows the same trend with similar magnitudes to the single point measurements (Fig. \ref{fgr4}d).

These LBT$^{3-}$:Tb$^{3+}$ results suggest that the coordination sphere of binding ligands around the Tb$^{3+}$ cation is stable in the highly anisotropic environment at the air-water interface. Further, the elastic transition with excess Tb$^{3+}$ for the negatively charged LBT$^{5-}$:Tb$^{3+}$ complex and the absence of such a transition for neutral LBT$^{3-}$:Tb$^{3+}$ complexes indicates that electrostatics play a key role in elastic structure formation. This concept is consistent with the hypothesis that neighboring LBT$^{5-}$:Tb$^{3+}$ complexes at the interface recruit Tb$^{3+}$. It is not clear, however, whether the elastic film formation relies on the presence of complexes with net negative charge, which are linked by Tb$^{3+}$ cations, or whether it depends more subtly on the location of the charged moieties in the bound complex.  

\subsection{Interfacial rheology of LBT$^{3-}$:Ca$^{2+}$ and RR-LBT$^{3-}$:Tb$^{3+}$ Complexes}
\begin{figure}[h]
 \centering
 \includegraphics[height=9.5cm]{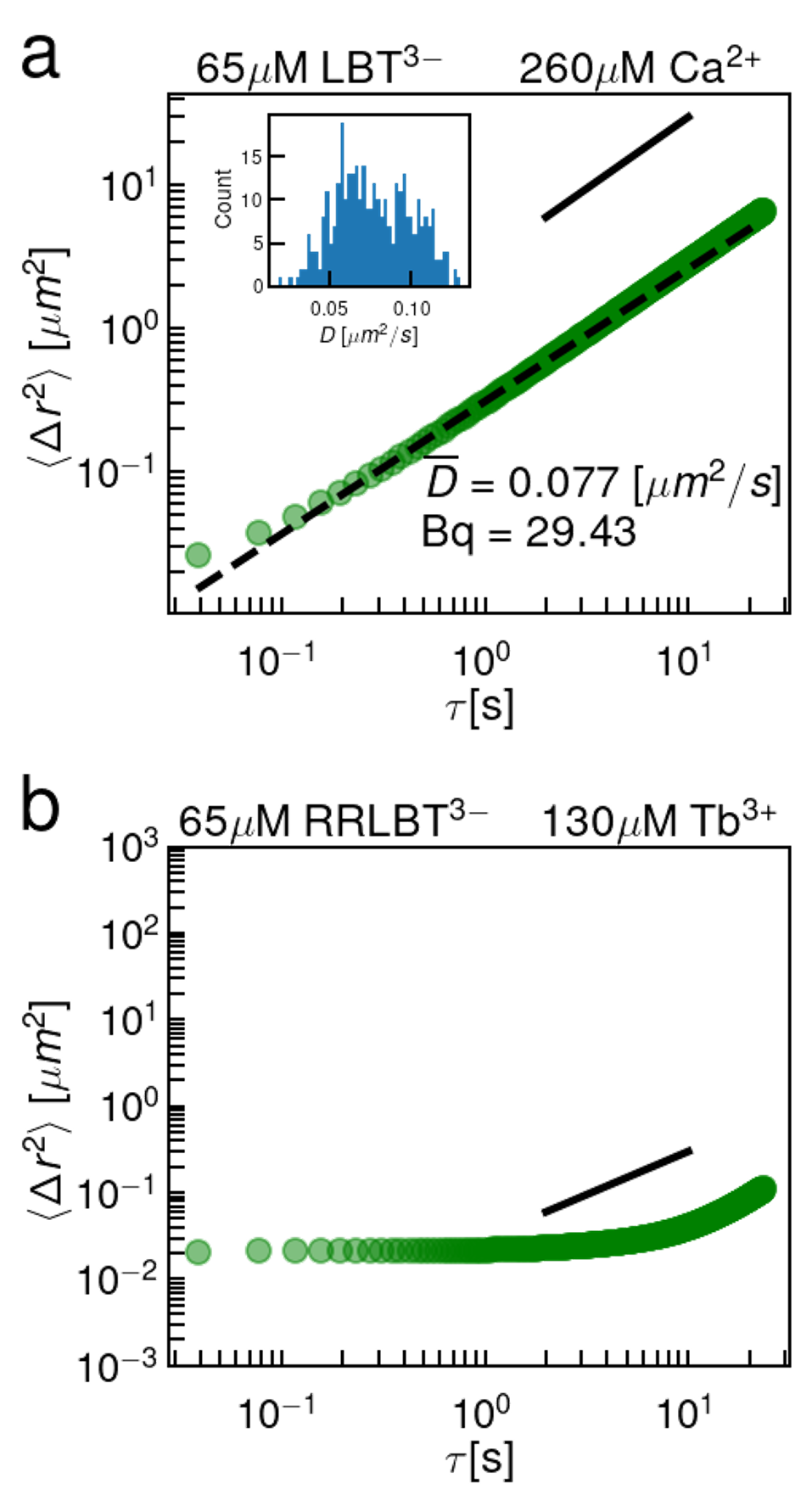}
 \caption{\textbf{Ligands outside of the binding loop are required for chelation and elastic film formation.} (a) EMSD of 65 $\mu$M LBT$^{3-}$ and 260 $\mu$M Ca$^{2+}$. Inset: the distribution of particle diffusivities showing a unimodal distribution of diffusivities.  (b) EMSD of 65 $\mu$M RR-LBT$^{3-}$ and 260 $\mu$M Tb$^{3+}$. The black dashed lines are fits of the form MSD = 4D$\tau$. The solid black lines are guides to show a power law with an exponent of 1.}
 \label{fgr5}
\end{figure}

To address this question, we have performed two sets of single point microrheology experiments. First, we exploit the ability of LBT$^{3-}$ to bind Ca$^{2+}$ to form a complex with a net negative charge.  Ca$^{2+}$ is an acceptable surrogate for Tb$^{3+}$ because they are of similar size and the binding loop was derived from a Ca$^{2+}$ binding protein.\cite{franz2003} The results (Fig. \ref{fgr5}a) show that the interface remains purely viscous in the presence of excess Ca$^{2+}$, i.e. at a Ca$^{2+}$ to LBT$^{3-}$ concentration ratio of 4.0. This result shows clearly that excess charge on the complex is not sufficient to promote network formation and suggests that a more subtle effect is at play. We hypothesize that elastic films do not form because LBT$^{3-}$:Ca$^{2+}$ complexes have no anionic ligands that are presented outside of the coordination sphere.  Those ligands within the sphere are sterically shielded from chelating with cations. Thus, while the LBT$^{3-}$:Ca$^{2+}$ complex has a net negative charge, the charged ligands are sterically hindered from chelating excess cations. To further probe the importance of ligands outside of the coordination sphere, another mutant, RR-LBT$^{3-}$, was prepared, in which two positively charged arginine residues have been appended to the N-terminus of LBT$^{5-}$ and the C-terminus is amidated. Based on the $K_D$ of RR-LBT$^{3-}$, shown in Fig. \ref{fgr1}, we infer that this peptide binds Tb$^{3+}$ similarly.  This mutant is designed to form net neutral complexes with Tb$^{3+}$. However, like LBT$^{5-}$, RR-LBT$^{3-}$ presents a carboxylate outside of the coordination sphere at the D11 position. Figure \ref{fgr5}b shows that despite the RR-LBT$^{3-}$:Tb$^{3+}$ complex having a neutral charge, adsorbed RR-LBT$^{3-}$:Tb$^{3+}$ complexes form thin elastic films in the presence of excess Tb$^{3+}$. 

These data indicate that, even in the absence of net negative charge on the complex, the negatively charged D11 residue is locally attractive and sterically accessible to the trivalent cation. Furthermore, these results indicate that the two positively charged arginines outside of the binding pocket do not repel the additional cation, suggesting “patchy colloid”-like interactions among the exposed charged sites on the peptide at the interface. The bulk Debye length for these solutions is around 1 nm. In the air-water interface, where the dielectric constant is about half that of the bulk\cite{miller2019}, the Debye length is closer to 0.5 nm which is sufficiently small such that patchy charges on these complexes could be distinguishable. This indicates that net neutral complexes are not sufficient to deter non-specific binding and a more stringent design criterion, having no anionic ligands outside the coordination sphere, is required.

To interrogate the coordination sphere in the presence of strong cationic perturbants, the rheology of PEPS laden interfaces was also probed in the presence of Al$^{3+}$. We select the small Al$^{3+}$ cation for two reasons. Its small size implies that its charge is highly concentrated, with a radius about half that of Ln$^{3+}$.\cite{shannon1976} This relates to a charge density almost 10x larger. Furthermore, it is too small to form a coordination sphere with LBT$^{3-}$ due to steric hindrance of the backbone. The addition of Al$^{3+}$ to LBT solutions is highly disruptive. Large extended aggregates form in the interface and in the bulk (Fig. S9); such aggregated clumps are never present for Tb$^{3+}$ or Ca$^{2+}$. Interestingly, the addition of Al$^{3+}$ to solutions of LBT$^{3-}$:Tb$^{3+}$ does not lead to such aggregate formation, providing additional evidence that the coordination sphere of LBT$^{3-}$:Tb$^{3+}$ is intact in the bulk and at the interface. Addition of Al$^{3+}$ to solutions of LBT$^{5-}$:Tb$^{3+}$ results in aggregation, whereas addition of Al$^{3+}$ to solutions of RR-LBT$^{3-}$:Tb$^{3+}$ shows no aggregation (Vid. S1-3). These results indicate that PEPS should be designed to be charge neutral to prevent electrostatic denaturation from cations with large charge densities in bulk solution.

\subsection{Molecular Dynamics and Metadynamics Simulations}
\begin{figure}[h]
 \centering
 \includegraphics[height=9.5cm]{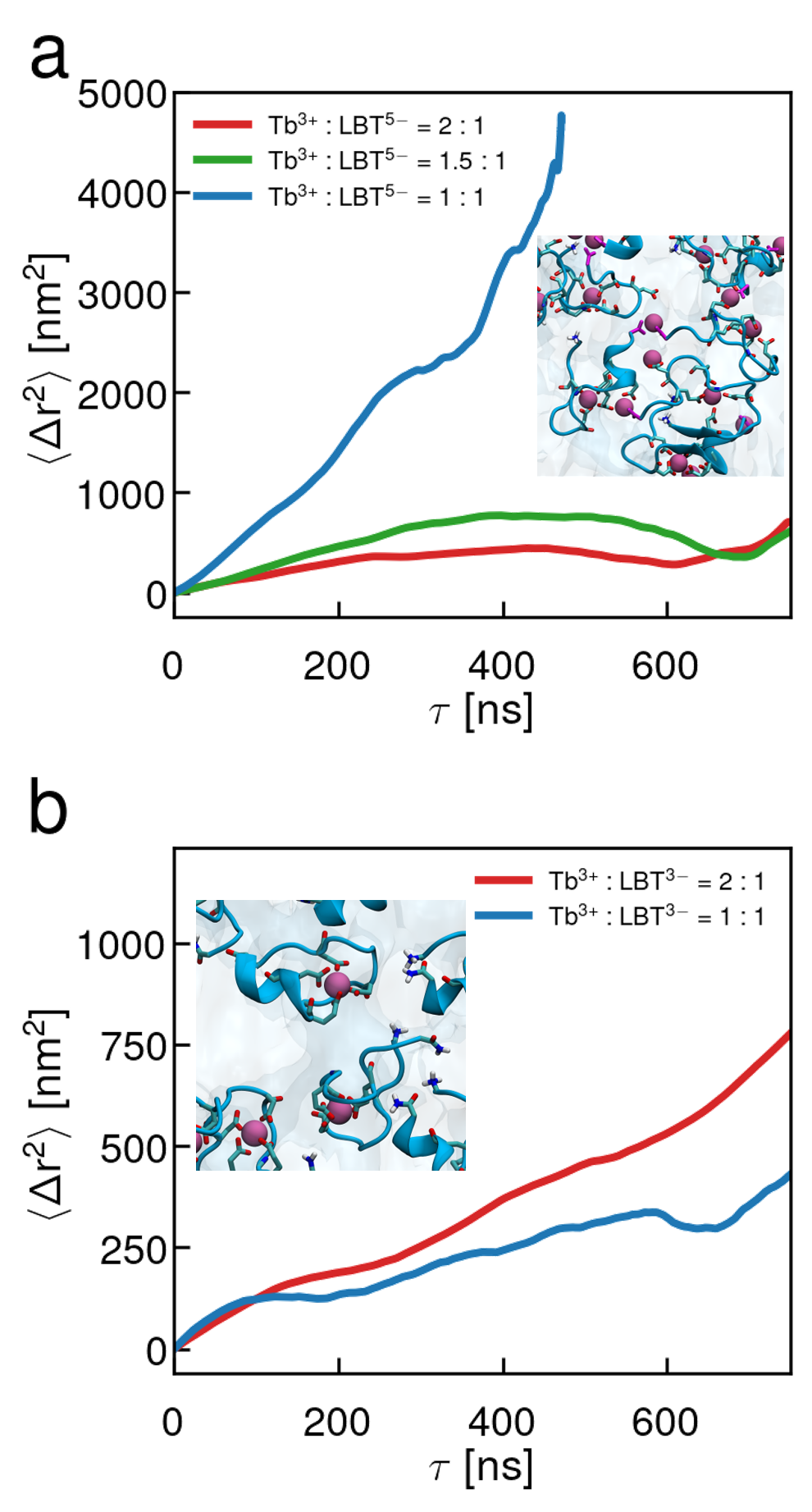}
 \caption{\textbf{Molecular dynamics simulations of PEPS in the air-aqueous interface show network formation occurring with LBT$^{5-}$ and excess Tb$^{3+}$ but not with LBT$^{3-}$ with excess Tb$^{3+}$.} EMSD of PEPS center of mass at the air-water interface for (a) LBT$^{5-}$ with (blue) 1:1 Tb$^{3+}$:LBT$^{5-}$; (green) 1.5:1 Tb$^{3+}$:LBT$^{5-}$; (red) 2:1 Tb$^{3+}$:LBT$^{5-}$ and (b) LBT$^{3-}$ with (blue) 1:1 Tb$^{3+}$:LBT$^{3-}$ and (red) 2:1 Tb$^{3+}$:LBT$^{3-}$. Insets show a representative snapshot of the interface at 2:1 Tb$^{3+}$:PEPS. }
 \label{fgr6}
\end{figure}

Our study of surface rheology suggests that the coordination spheres formed by the LBT-derived peptides do not denature at the interface, even in the presence of excess large ions like Tb$^{3+}$ or Ca$^{2+}$. Furthermore, the experimental results show that the presence of non-binding loop ligands is responsible for interfacial elasticity and network formation. To get insight into molecular conformations at the interface at various Tb$^{3+}$ to peptide ratios, all-atom MD simulations were performed on air-aqueous interfaces of solutions of PEPS and Tb$^{3+}$.  The results from the MD simulations corroborate the findings of the particle tracking microrheology, in that the displacements are highly attenuated in the presence of excess cations (Fig. \ref{fgr6}).

\begin{figure*}
 \centering
 \includegraphics[height=8.5cm]{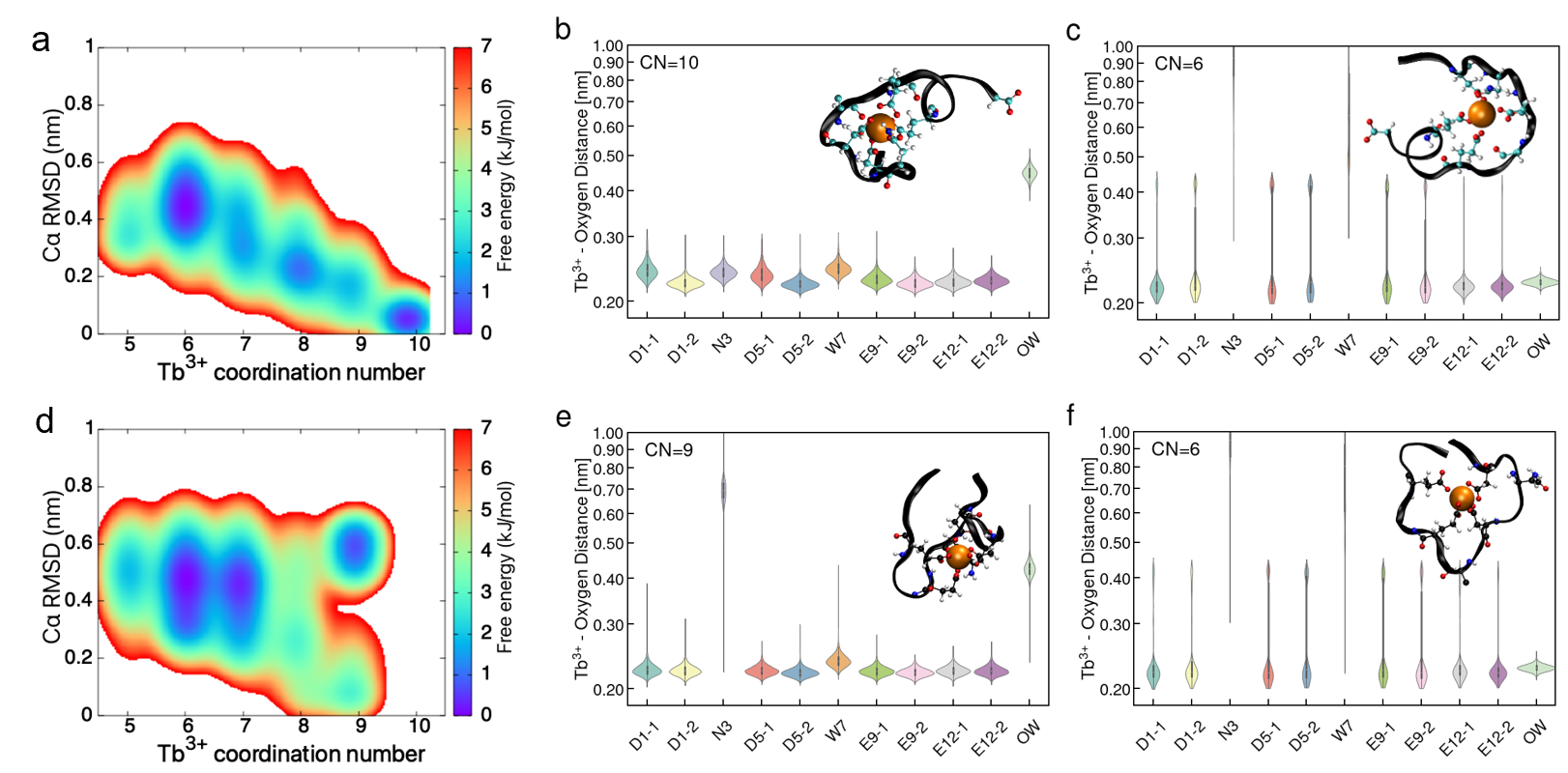}
 \caption{\textbf{Metadynamics simulation of interfacial PEPS structure shows most probable peptide conformations have all anionic binding loop ligands occupied by bound Ln$^{3+}$.} (a, d) Metadynamics simulation conformation energy landscape for LBT$^{5-}$ and LBT$^{3-}$. Distance between Tb$^{3+}$ and the residue oxygen for the prominent energy wells of (b, c) LBT$^{5-}$ and (e, f) LBT$^{3-}$. }
 \label{fgr7}
\end{figure*}
Figure \ref{fgr6} shows that LBT$^{5-}$ undergoes network formation as Tb$^{3+}$ concentration increases past the stoichiometric ratio. The EMSD data is subtle, but for larger than stoichiometric ratios of Tb$^{3+}$ (Fig. \ref{fgr6}a) a plateau can be seen at later lag times around 200 ns. We attribute this late onset plateau to a characteristic length of the elastic network that forms. As such, PEPS move diffusively until they have moved far enough to strain the elastic network and then the PEPS EMSD shows a “caged” behavior. It appears that this characteristic length decreases with increasing Tb$^{3+}$ concentration i.e. the plateaus occur at slightly lower MSD when greater excess Tb$^{3+}$ is present indicating that more ions may crosslink the network more densely. In the simulations for LBT$^{3-}$, the EMSDs are increasing during the entire duration of the simulation. This implies that the interface is viscous even when excess Tb$^{3+}$ is present (Fig. \ref{fgr6}b).  Careful inspection of the inset MD snapshots reveals how excess ions are present in the interfacial layer for LBT$^{5-}$ and crosslink multiple LBT$^{5-}$ peptides together. While similar results are found for RRLBT$^3-$, this behavior is not seen for interfacial layers of LBT$^{3-}$. Additional snapshots showing the sideview of the interface are shown in Fig. S7b-d which show how excess Tb$^{3+}$ localizes at the interface for LBT$^{5-}$ but not for LBT$^{3-}$. Lastly, the PEPS in the interfacial layer appear to have intact binding loops throughout the duration of the simulation which suggests the binding loop is robust, again corroborating the findings of microrheology. Animations of the simulations can be found in the supplementary information (Vid. S4-6).

In addition to traditional MD, metadynamics simulations were performed to access PEPS conformations that may not be available to normal MD simulation. Figure \ref{fgr7} shows the results of the metadynamics simulation. Here, we have performed well-tempered metadynamics simulations to further probe the molecular conformation of PEPS:Tb$^{3+}$ binding complex at the air-water interface. Metadynamics simulations of the interface identified 5 to 6 PEPS:Tb$^{3+}$ binding conformations with small free energy differences ($\Delta\Delta G \leq 1$ kcal/mol), revealing the hidden complexity of the PEPS:Tb$^{3+}$ binding complex conformations (Fig. \ref{fgr7}a, d). For instance, LBT$^{5-}$ tends to sample three conformations with relatively higher probability (thus lower free energy) where Tb$^{3+}$ is coordinated with 10, 8, and 6 oxygen atoms from LBT$^{5-}$ respectively. (See Fig. S10 for the ranking of basin free energy value of each binding conformation). By comparison, LBT$^{3-}$ tends to have 9, 7, and 6 oxygen atoms coordinate with Tb$^{3+}$ in the binding loop, suggesting that LBT$^{3-}$ may bind more weakly than LBT$^{5-}$ with Tb$^{3+}$ at the interface.  We report the residue oxygen-Tb$^{3+}$ distance distances in violin plots (Fig. \ref{fgr7}b, c, e, f), for the most highly coordinated and probable state and the least coordinated and probable state: 10 and 6 coordinate for LBT$^{5-}$ and 9 and 6 coordinate for LBT$^{3-}$. The intermediate coordination states (8 coordinate for LBT$^{5-}$ and 7 coordinate for LBT$^{3-}$) can be found in Fig. S11. Analysis of the residue oxygen-Tb$^{3+}$ distance shows that in the case of both LBT$^{5-}$  and LBT$^{3-}$  all carboxylate ligands are bound to the Tb$^{3+}$ cation in the interface even with lower coordinate structures. This implies that the elasticity seen with LBT$^{5-}$ comes from the non-binding loop ligands. For LBT$^{3-}$ no elasticity is observed and metadynamics reports that all anionic ligands are interacting with a Tb$^{3+}$ in the binding loop. Since these anionic ligands are already in close contact with a cation inside the binding loop, they cannot be presented in such a way to crosslink with extra cations outside the binding loop.

\section{Conclusion}
This study demonstrates that microrheology by particle tracking can be used to infer molecular conformations of metal binding peptides that adsorb to air-water interfaces.  The inferred molecular structures were further examined and confirmed by MD and metadynamics simulations. Recalling Figure \ref{fgr1}, we are able to categorize our PEPS into each of the scenarios. Scenario I represents PEPS that maintain their binding loop in the air-aqueous interface even in conditions with excess Tb$^{3+}$. We demonstrate that LBT$^{3-}$ fits this scenario due to its purely viscous rheology, failure to denature upon addition of Al$^{3+}$ to PEPS that have already bound Tb$^{3+}$, and prior study that shows 1:1 binding. In addition, metadynamics simulation shows the lowest free energy conformations of LBT$^{3-}$ have all binding loop ligands facing inward and interacting with the bound Tb$^{3+}$ in the interface. In the presence of excess Ca$^{2+}$, where the Ca$^{2+}$:LBT$^{3-}$ complex carried negative charge, electrostatic interaction with excess Ca$^{2+}$ was not strong enough to denature the coordination sphere and lead to elastic network formation. We suggest that the subtle detail of this PEPS which mitigates non-specific binding is the lack of anionic ligands outside of the binding loop and consider that a design criterion for future PEPS development. The other PEPS studied fall in Scenario II. Both LBT$^{5-}$ and RR-LBT$^{3-}$ formed elastic networks at the air-aqueous interface of solutions containing excess Tb$^{3+}$. Both of these PEPS contain non-binding loop ligands, notably D11 in both and also the C-terminal carboxylate in LBT$^{5-}$. While these PEPS form an elastic network, it should be emphasized that their binding loop is robust in the interface and likely stays coordinated. This is not only confirmed with careful mutation and ion selection (LBT$^{3-}$ with Ca$^{2+}$), but also through interfacial MD and metadynamics studies which show the binding loop ligands are always interacting with the bound Tb$^{3+}$ in the lowest energy configurations. Studies of PEPS with Al$^{3+}$, a small and densely charged cation prone to denaturing PEPS, show that denatured PEPS form large extended aggregates which are never seen with LBT$^{5-}$ and RR-LBT$^{3-}$ in the presence of Tb$^{3+}$. This evidence suggests that Scenario III does not happen and that the binding loops are robust in the interface. Interfacial elastic network formation likely occurs through interactions of Tb$^{3+}$ with non-binding loop ligands that face outwards from the coordination sphere. The Al$^{3+}$ denaturation studies show another critical design criterion: PEPS must be net neutral to avoid non-Ln$^{3+}$ induced denaturation and aggregation. 

Our work demonstrates that PEPS can be designed to carry Ln$^{3+}$ to the air-aqueous interface and maintain the selective 1:1 binding loop structure in the interface even under perturbation from the anisotropic environment and excess ions. There are two criteria that should be met: net neutrality of the complex and no anionic ligands outside of the binding loop. In this way, PEPS can bind and retain REE cations in their binding loop at the interface without non-selective Coulombic interactions to provide a means of green selective separation of REEs.

\section*{Author Contributions}
KJS, IJD, CM, RST, CFN, RR, EJP designed the study, supervised research, and aided in interpretation of results. SAC performed all rheology experiments, analysis and interpretation with guidance from JD and MM. FJA performed and analyzed the molecular dynamics simulations, MOdlC participated in their interpretation. YW performed metadynamics simulations; RR aided in their design and interpretation. LEOM and JGM provided guidance in sample preparation. LEOM interpreted results. JGM carried out binding assays. The manuscript was drafted by SAC and reviewed and edited by all authors.

\section*{Conflicts of interest}
There are no conflicts to declare.

\section*{Acknowledgements}
We acknowledge Dr. Nicholas. G. Chisholm's assistance with constructing theoretical interfacial Stokeslets.

\noindent This research was made possible by a grant from the Department of Energy (DOE-BES Award DE-SC0022240). We thank the Vagelos Insitute for Energy Science and Technology Graduate Fellowship for their support of one of the authors.

\bibliography{refs} 

\providecommand*{\mcitethebibliography}{\thebibliography}
\csname @ifundefined\endcsname{endmcitethebibliography}
{\let\endmcitethebibliography\endthebibliography}{}
\begin{mcitethebibliography}{43}
\providecommand*{\natexlab}[1]{#1}
\providecommand*{\mciteSetBstSublistMode}[1]{}
\providecommand*{\mciteSetBstMaxWidthForm}[2]{}
\providecommand*{\mciteBstWouldAddEndPuncttrue}
  {\def\EndOfBibitem{\unskip.}}
\providecommand*{\mciteBstWouldAddEndPunctfalse}
  {\let\EndOfBibitem\relax}
\providecommand*{\mciteSetBstMidEndSepPunct}[3]{}
\providecommand*{\mciteSetBstSublistLabelBeginEnd}[3]{}
\providecommand*{\EndOfBibitem}{}
\mciteSetBstSublistMode{f}
\mciteSetBstMaxWidthForm{subitem}
{(\emph{\alph{mcitesubitemcount}})}
\mciteSetBstSublistLabelBeginEnd{\mcitemaxwidthsubitemform\space}
{\relax}{\relax}

\bibitem[Van~Gosen \emph{et~al.}(2017)Van~Gosen, Verplanck, Seal~Ii, Long, and Gambogi]{vangosen2017}
B.~S. Van~Gosen, P.~L. Verplanck, R.~R. Seal~Ii, K.~R. Long and J.~Gambogi, \emph{Rare-earth elements}, U.S. Geological Survey Professional Paper Report 1802, 2017\relax
\mciteBstWouldAddEndPuncttrue
\mciteSetBstMidEndSepPunct{\mcitedefaultmidpunct}
{\mcitedefaultendpunct}{\mcitedefaultseppunct}\relax
\EndOfBibitem
\bibitem[Long(2011)]{long2011}
K.~Long, \emph{The future of rare earth elements—will these high-tech industry elements continue in short supply?}, U.S. Geological Survey Open-File Report 1189, 2011\relax
\mciteBstWouldAddEndPuncttrue
\mciteSetBstMidEndSepPunct{\mcitedefaultmidpunct}
{\mcitedefaultendpunct}{\mcitedefaultseppunct}\relax
\EndOfBibitem
\bibitem[Goonan(2011)]{goonan2011}
T.~G. Goonan, \emph{Rare Earth Elements—End Use and Recyclability}, U.S. Geological Survey Scientific Investigations Report 5094, 2011\relax
\mciteBstWouldAddEndPuncttrue
\mciteSetBstMidEndSepPunct{\mcitedefaultmidpunct}
{\mcitedefaultendpunct}{\mcitedefaultseppunct}\relax
\EndOfBibitem
\bibitem[Opare \emph{et~al.}(2021)Opare, Struhs, and Mirkouei]{opare2021}
E.~O. Opare, E.~Struhs and A.~Mirkouei, \emph{Renewable \& Sustainable Energy Reviews}, 2021, \textbf{143}, 110917\relax
\mciteBstWouldAddEndPuncttrue
\mciteSetBstMidEndSepPunct{\mcitedefaultmidpunct}
{\mcitedefaultendpunct}{\mcitedefaultseppunct}\relax
\EndOfBibitem
\bibitem[Liu and Chen(2021)]{liu2021}
T.~C. Liu and J.~Chen, \emph{Separation and Purification Technology}, 2021, \textbf{276}, 119263\relax
\mciteBstWouldAddEndPuncttrue
\mciteSetBstMidEndSepPunct{\mcitedefaultmidpunct}
{\mcitedefaultendpunct}{\mcitedefaultseppunct}\relax
\EndOfBibitem
\bibitem[Cheisson and Schelter(2019)]{cheisson2019}
T.~Cheisson and E.~J. Schelter, \emph{Science}, 2019, \textbf{363}, 489--493\relax
\mciteBstWouldAddEndPuncttrue
\mciteSetBstMidEndSepPunct{\mcitedefaultmidpunct}
{\mcitedefaultendpunct}{\mcitedefaultseppunct}\relax
\EndOfBibitem
\bibitem[Xu \emph{et~al.}(2019)Xu, Su, and Renner]{xu2019}
M.~Y. Xu, Z.~H. Su and J.~N. Renner, \emph{Peptide Science}, 2019, \textbf{111}, e24133\relax
\mciteBstWouldAddEndPuncttrue
\mciteSetBstMidEndSepPunct{\mcitedefaultmidpunct}
{\mcitedefaultendpunct}{\mcitedefaultseppunct}\relax
\EndOfBibitem
\bibitem[Su \emph{et~al.}(2021)Su, Hostert, and Renner]{su2021}
Z.~H. Su, J.~D. Hostert and J.~N. Renner, \emph{Acs Es\&T Water}, 2021, \textbf{1}, 58--67\relax
\mciteBstWouldAddEndPuncttrue
\mciteSetBstMidEndSepPunct{\mcitedefaultmidpunct}
{\mcitedefaultendpunct}{\mcitedefaultseppunct}\relax
\EndOfBibitem
\bibitem[Hostert \emph{et~al.}(2023)Hostert, Sepesy, Duval, and Renner]{hostert2023}
J.~D. Hostert, M.~R. Sepesy, C.~E. Duval and J.~N. Renner, \emph{Soft Matter}, 2023, \textbf{19}, 2823--2831\relax
\mciteBstWouldAddEndPuncttrue
\mciteSetBstMidEndSepPunct{\mcitedefaultmidpunct}
{\mcitedefaultendpunct}{\mcitedefaultseppunct}\relax
\EndOfBibitem
\bibitem[Martin \emph{et~al.}(2005)Martin, Sculimbrene, Nitz, and Imperiali]{martin2005}
L.~L. Martin, B.~R. Sculimbrene, M.~Nitz and B.~Imperiali, \emph{Qsar \& Combinatorial Science}, 2005, \textbf{24}, 1149--1157\relax
\mciteBstWouldAddEndPuncttrue
\mciteSetBstMidEndSepPunct{\mcitedefaultmidpunct}
{\mcitedefaultendpunct}{\mcitedefaultseppunct}\relax
\EndOfBibitem
\bibitem[Franz \emph{et~al.}(2003)Franz, Nitz, and Imperiali]{franz2003}
K.~J. Franz, M.~Nitz and B.~Imperiali, \emph{Chembiochem}, 2003, \textbf{4}, 265--271\relax
\mciteBstWouldAddEndPuncttrue
\mciteSetBstMidEndSepPunct{\mcitedefaultmidpunct}
{\mcitedefaultendpunct}{\mcitedefaultseppunct}\relax
\EndOfBibitem
\bibitem[Nitz \emph{et~al.}(2004)Nitz, Sherawat, Franz, Peisach, Allen, and Imperiali]{nitz2004}
M.~Nitz, M.~Sherawat, K.~J. Franz, E.~Peisach, K.~N. Allen and B.~Imperiali, \emph{Angewandte Chemie-International Edition}, 2004, \textbf{43}, 3682--3685\relax
\mciteBstWouldAddEndPuncttrue
\mciteSetBstMidEndSepPunct{\mcitedefaultmidpunct}
{\mcitedefaultendpunct}{\mcitedefaultseppunct}\relax
\EndOfBibitem
\bibitem[Ortuno~Macias \emph{et~al.}(unpublished work)Ortuno~Macias, Jimenez-Angeles, Marmorstein, Wang, Crane, KT, Sun, Sapkota, Hummingbird, Jung, Qiao, Lee, Dmochowski, Messinger, Schlossman, Radhakrishnan, Petersson, Olvera de~la Cruz, Bu, Bera, Lin, Tu, Stebe, and Maldarelli]{luis2024}
L.~Ortuno~Macias, F.~Jimenez-Angeles, J.~G. Marmorstein, Y.~Wang, S.~A. Crane, S.~KT, P.~Sun, B.~Sapkota, E.~Hummingbird, W.~Jung, B.~Qiao, D.~Lee, I.~J. Dmochowski, R.~Messinger, M.~L. Schlossman, R.~Radhakrishnan, E.~J. Petersson, M.~Olvera de~la Cruz, W.~Bu, M.~Bera, B.~Lin, R.~Tu, K.~Stebe and C.~Maldarelli, unpublished work\relax
\mciteBstWouldAddEndPuncttrue
\mciteSetBstMidEndSepPunct{\mcitedefaultmidpunct}
{\mcitedefaultendpunct}{\mcitedefaultseppunct}\relax
\EndOfBibitem
\bibitem[Pace \emph{et~al.}(1995)Pace, Vajdos, Fee, Grimsley, and Gray]{pace1995}
C.~N. Pace, F.~Vajdos, L.~Fee, G.~Grimsley and T.~Gray, \emph{Protein Science}, 1995, \textbf{4}, 2411--2423\relax
\mciteBstWouldAddEndPuncttrue
\mciteSetBstMidEndSepPunct{\mcitedefaultmidpunct}
{\mcitedefaultendpunct}{\mcitedefaultseppunct}\relax
\EndOfBibitem
\bibitem[Lee \emph{et~al.}(2010)Lee, Reich, Stebe, and Leheny]{lee2010}
M.~H. Lee, D.~H. Reich, K.~J. Stebe and R.~L. Leheny, \emph{Langmuir}, 2010, \textbf{26}, 2650--2658\relax
\mciteBstWouldAddEndPuncttrue
\mciteSetBstMidEndSepPunct{\mcitedefaultmidpunct}
{\mcitedefaultendpunct}{\mcitedefaultseppunct}\relax
\EndOfBibitem
\bibitem[Deng \emph{et~al.}(2020)Deng, Molaei, Chisholm, and Stebe]{deng2020}
J.~Y. Deng, M.~Molaei, N.~G. Chisholm and K.~J. Stebe, \emph{Langmuir}, 2020, \textbf{36}, 6888--6902\relax
\mciteBstWouldAddEndPuncttrue
\mciteSetBstMidEndSepPunct{\mcitedefaultmidpunct}
{\mcitedefaultendpunct}{\mcitedefaultseppunct}\relax
\EndOfBibitem
\bibitem[Gharbi \emph{et~al.}(2011)Gharbi, Nobili, In, Prévot, Galatola, Fournier, and Blanc]{gharbi2011}
M.~A. Gharbi, M.~Nobili, M.~In, G.~Prévot, P.~Galatola, J.~B. Fournier and C.~Blanc, \emph{Soft Matter}, 2011, \textbf{7}, 1467--1471\relax
\mciteBstWouldAddEndPuncttrue
\mciteSetBstMidEndSepPunct{\mcitedefaultmidpunct}
{\mcitedefaultendpunct}{\mcitedefaultseppunct}\relax
\EndOfBibitem
\bibitem[Crocker and Grier(1996)]{crocker1996}
J.~C. Crocker and D.~G. Grier, \emph{Journal of Colloid and Interface Science}, 1996, \textbf{179}, 298--310\relax
\mciteBstWouldAddEndPuncttrue
\mciteSetBstMidEndSepPunct{\mcitedefaultmidpunct}
{\mcitedefaultendpunct}{\mcitedefaultseppunct}\relax
\EndOfBibitem
\bibitem[Allan \emph{et~al.}(2024)Allan, Caswell, Keim, van~der Wel, and Verweij]{allan_2024_10696534}
D.~B. Allan, T.~Caswell, N.~C. Keim, C.~M. van~der Wel and R.~W. Verweij, \emph{soft-matter/trackpy: v0.6.2}, 2024, \url{https://doi.org/10.5281/zenodo.10696534}\relax
\mciteBstWouldAddEndPuncttrue
\mciteSetBstMidEndSepPunct{\mcitedefaultmidpunct}
{\mcitedefaultendpunct}{\mcitedefaultseppunct}\relax
\EndOfBibitem
\bibitem[Saffman and Delbruck(1975)]{saffman1975}
P.~G. Saffman and M.~Delbruck, \emph{Proceedings of the National Academy of Sciences of the United States of America}, 1975, \textbf{72}, 3111--3113\relax
\mciteBstWouldAddEndPuncttrue
\mciteSetBstMidEndSepPunct{\mcitedefaultmidpunct}
{\mcitedefaultendpunct}{\mcitedefaultseppunct}\relax
\EndOfBibitem
\bibitem[Saffman(1976)]{saffman1976}
P.~G. Saffman, \emph{Journal of Fluid Mechanics}, 1976, \textbf{73}, 593--602\relax
\mciteBstWouldAddEndPuncttrue
\mciteSetBstMidEndSepPunct{\mcitedefaultmidpunct}
{\mcitedefaultendpunct}{\mcitedefaultseppunct}\relax
\EndOfBibitem
\bibitem[Hughes \emph{et~al.}(1981)Hughes, Pailthorpe, and White]{hughes1981}
B.~D. Hughes, B.~A. Pailthorpe and L.~R. White, \emph{Journal of Fluid Mechanics}, 1981, \textbf{110}, 349--372\relax
\mciteBstWouldAddEndPuncttrue
\mciteSetBstMidEndSepPunct{\mcitedefaultmidpunct}
{\mcitedefaultendpunct}{\mcitedefaultseppunct}\relax
\EndOfBibitem
\bibitem[Fischer \emph{et~al.}(2006)Fischer, Dhar, and Heinig]{fischer2006}
T.~M. Fischer, P.~Dhar and P.~Heinig, \emph{Journal of Fluid Mechanics}, 2006, \textbf{558}, 451--475\relax
\mciteBstWouldAddEndPuncttrue
\mciteSetBstMidEndSepPunct{\mcitedefaultmidpunct}
{\mcitedefaultendpunct}{\mcitedefaultseppunct}\relax
\EndOfBibitem
\bibitem[Mason and Weitz(1995)]{mason1995}
T.~G. Mason and D.~A. Weitz, \emph{Physical Review Letters}, 1995, \textbf{74}, 1250--1253\relax
\mciteBstWouldAddEndPuncttrue
\mciteSetBstMidEndSepPunct{\mcitedefaultmidpunct}
{\mcitedefaultendpunct}{\mcitedefaultseppunct}\relax
\EndOfBibitem
\bibitem[Dasgupta \emph{et~al.}(2002)Dasgupta, Tee, Crocker, Frisken, and Weitz]{dasgupta2002}
B.~R. Dasgupta, S.~Y. Tee, J.~C. Crocker, B.~J. Frisken and D.~A. Weitz, \emph{Physical Review E}, 2002, \textbf{65}, \relax
\mciteBstWouldAddEndPuncttrue
\mciteSetBstMidEndSepPunct{\mcitedefaultmidpunct}
{\mcitedefaultendpunct}{\mcitedefaultseppunct}\relax
\EndOfBibitem
\bibitem[Molaei \emph{et~al.}(2021)Molaei, Chisholm, Deng, Crocker, and Stebe]{molaei2021}
M.~Molaei, N.~G. Chisholm, J.~Y. Deng, J.~C. Crocker and K.~J. Stebe, \emph{Physical Review Letters}, 2021, \textbf{126}, 228003\relax
\mciteBstWouldAddEndPuncttrue
\mciteSetBstMidEndSepPunct{\mcitedefaultmidpunct}
{\mcitedefaultendpunct}{\mcitedefaultseppunct}\relax
\EndOfBibitem
\bibitem[Chisholm and Stebe(2021)]{chisholm2021}
N.~G. Chisholm and K.~J. Stebe, \emph{Journal of Fluid Mechanics}, 2021, \textbf{914}, A29\relax
\mciteBstWouldAddEndPuncttrue
\mciteSetBstMidEndSepPunct{\mcitedefaultmidpunct}
{\mcitedefaultendpunct}{\mcitedefaultseppunct}\relax
\EndOfBibitem
\bibitem[Stone and Ajdari(1998)]{stone1998}
H.~A. Stone and A.~Ajdari, \emph{Journal of Fluid Mechanics}, 1998, \textbf{369}, 151--173\relax
\mciteBstWouldAddEndPuncttrue
\mciteSetBstMidEndSepPunct{\mcitedefaultmidpunct}
{\mcitedefaultendpunct}{\mcitedefaultseppunct}\relax
\EndOfBibitem
\bibitem[Prasad \emph{et~al.}(2006)Prasad, Koehler, and Weeks]{prasad2006}
V.~Prasad, S.~A. Koehler and E.~R. Weeks, \emph{Physical Review Letters}, 2006, \textbf{97}, 176001\relax
\mciteBstWouldAddEndPuncttrue
\mciteSetBstMidEndSepPunct{\mcitedefaultmidpunct}
{\mcitedefaultendpunct}{\mcitedefaultseppunct}\relax
\EndOfBibitem
\bibitem[Rotenberg \emph{et~al.}(1983)Rotenberg, Boruvka, and Neumann]{rotenberg1983}
Y.~Rotenberg, L.~Boruvka and A.~W. Neumann, \emph{Journal of Colloid and Interface Science}, 1983, \textbf{93}, 169--183\relax
\mciteBstWouldAddEndPuncttrue
\mciteSetBstMidEndSepPunct{\mcitedefaultmidpunct}
{\mcitedefaultendpunct}{\mcitedefaultseppunct}\relax
\EndOfBibitem
\bibitem[Jiménez-Angeles \emph{et~al.}(2019)Jiménez-Angeles, Kwon, Sadman, Wu, Shull, and de~la Cruz]{jimenez2019}
F.~Jiménez-Angeles, H.~K. Kwon, K.~Sadman, T.~Wu, K.~R. Shull and M.~O. de~la Cruz, \emph{Acs Central Science}, 2019, \textbf{5}, 688--699\relax
\mciteBstWouldAddEndPuncttrue
\mciteSetBstMidEndSepPunct{\mcitedefaultmidpunct}
{\mcitedefaultendpunct}{\mcitedefaultseppunct}\relax
\EndOfBibitem
\bibitem[Páll \emph{et~al.}(2020)Páll, Zhmurov, Bauer, Abraham, Lundborg, Gray, Hess, and Lindahl]{pall2020}
S.~Páll, A.~Zhmurov, P.~Bauer, M.~Abraham, M.~Lundborg, A.~Gray, B.~Hess and E.~Lindahl, \emph{Journal of Chemical Physics}, 2020, \textbf{153}, 134110\relax
\mciteBstWouldAddEndPuncttrue
\mciteSetBstMidEndSepPunct{\mcitedefaultmidpunct}
{\mcitedefaultendpunct}{\mcitedefaultseppunct}\relax
\EndOfBibitem
\bibitem[Hess \emph{et~al.}(2008)Hess, Kutzner, van~der Spoel, and Lindahl]{hess2008}
B.~Hess, C.~Kutzner, D.~van~der Spoel and E.~Lindahl, \emph{Journal of Chemical Theory and Computation}, 2008, \textbf{4}, 435--447\relax
\mciteBstWouldAddEndPuncttrue
\mciteSetBstMidEndSepPunct{\mcitedefaultmidpunct}
{\mcitedefaultendpunct}{\mcitedefaultseppunct}\relax
\EndOfBibitem
\bibitem[Abraham \emph{et~al.}(2015)Abraham, Murtola, Schulz, Páll, Smith, Hess, and Lindahl]{abraham2015}
M.~J. Abraham, T.~Murtola, R.~Schulz, S.~Páll, J.~C. Smith, B.~Hess and E.~Lindahl, \emph{SoftwareX}, 2015, \textbf{1}, 19--25\relax
\mciteBstWouldAddEndPuncttrue
\mciteSetBstMidEndSepPunct{\mcitedefaultmidpunct}
{\mcitedefaultendpunct}{\mcitedefaultseppunct}\relax
\EndOfBibitem
\bibitem[Brooks \emph{et~al.}(2009)Brooks, Brooks, Mackerell, Nilsson, Petrella, Roux, Won, Archontis, Bartels, Boresch, Caflisch, Caves, Cui, Dinner, Feig, Fischer, Gao, Hodoscek, Im, Kuczera, Lazaridis, Ma, Ovchinnikov, Paci, Pastor, Post, Pu, Schaefer, Tidor, Venable, Woodcock, Wu, Yang, York, and Karplus]{brooks2009}
B.~R. Brooks, C.~L. Brooks, A.~D. Mackerell, L.~Nilsson, R.~J. Petrella, B.~Roux, Y.~Won, G.~Archontis, C.~Bartels, S.~Boresch, A.~Caflisch, L.~Caves, Q.~Cui, A.~R. Dinner, M.~Feig, S.~Fischer, J.~Gao, M.~Hodoscek, W.~Im, K.~Kuczera, T.~Lazaridis, J.~Ma, V.~Ovchinnikov, E.~Paci, R.~W. Pastor, C.~B. Post, J.~Z. Pu, M.~Schaefer, B.~Tidor, R.~M. Venable, H.~L. Woodcock, X.~Wu, W.~Yang, D.~M. York and M.~Karplus, \emph{Journal of Computational Chemistry}, 2009, \textbf{30}, 1545--1614\relax
\mciteBstWouldAddEndPuncttrue
\mciteSetBstMidEndSepPunct{\mcitedefaultmidpunct}
{\mcitedefaultendpunct}{\mcitedefaultseppunct}\relax
\EndOfBibitem
\bibitem[Nose(1984)]{nose1984}
S.~Nose, \emph{Journal of Chemical Physics}, 1984, \textbf{81}, 511--519\relax
\mciteBstWouldAddEndPuncttrue
\mciteSetBstMidEndSepPunct{\mcitedefaultmidpunct}
{\mcitedefaultendpunct}{\mcitedefaultseppunct}\relax
\EndOfBibitem
\bibitem[Hoover(1985)]{hoover1985}
W.~G. Hoover, \emph{Physical Review A}, 1985, \textbf{31}, 1695--1697\relax
\mciteBstWouldAddEndPuncttrue
\mciteSetBstMidEndSepPunct{\mcitedefaultmidpunct}
{\mcitedefaultendpunct}{\mcitedefaultseppunct}\relax
\EndOfBibitem
\bibitem[Bonomi \emph{et~al.}(2009)Bonomi, Branduardi, Bussi, Camilloni, Provasi, Raiteri, Donadio, Marinelli, Pietrucci, Broglia, and Parrinello]{bonomi2009}
M.~Bonomi, D.~Branduardi, G.~Bussi, C.~Camilloni, D.~Provasi, P.~Raiteri, D.~Donadio, F.~Marinelli, F.~Pietrucci, R.~A. Broglia and M.~Parrinello, \emph{Computer Physics Communications}, 2009, \textbf{180}, 1961--1972\relax
\mciteBstWouldAddEndPuncttrue
\mciteSetBstMidEndSepPunct{\mcitedefaultmidpunct}
{\mcitedefaultendpunct}{\mcitedefaultseppunct}\relax
\EndOfBibitem
\bibitem[Krivov \emph{et~al.}(2009)Krivov, Shapovalov, and Dunbrack]{krivov2009}
G.~G. Krivov, M.~V. Shapovalov and R.~L. Dunbrack, \emph{Proteins-Structure Function and Bioinformatics}, 2009, \textbf{77}, 778--795\relax
\mciteBstWouldAddEndPuncttrue
\mciteSetBstMidEndSepPunct{\mcitedefaultmidpunct}
{\mcitedefaultendpunct}{\mcitedefaultseppunct}\relax
\EndOfBibitem
\bibitem[Bussi \emph{et~al.}(2007)Bussi, Donadio, and Parrinello]{bussi2007}
G.~Bussi, D.~Donadio and M.~Parrinello, \emph{Journal of Chemical Physics}, 2007, \textbf{126}, 014101\relax
\mciteBstWouldAddEndPuncttrue
\mciteSetBstMidEndSepPunct{\mcitedefaultmidpunct}
{\mcitedefaultendpunct}{\mcitedefaultseppunct}\relax
\EndOfBibitem
\bibitem[Crocker \emph{et~al.}(2000)Crocker, Valentine, Weeks, Gisler, Kaplan, Yodh, and Weitz]{crocker2000}
J.~C. Crocker, M.~T. Valentine, E.~R. Weeks, T.~Gisler, P.~D. Kaplan, A.~G. Yodh and D.~A. Weitz, \emph{Physical Review Letters}, 2000, \textbf{85}, 888--891\relax
\mciteBstWouldAddEndPuncttrue
\mciteSetBstMidEndSepPunct{\mcitedefaultmidpunct}
{\mcitedefaultendpunct}{\mcitedefaultseppunct}\relax
\EndOfBibitem
\bibitem[Miller \emph{et~al.}(2019)Miller, Liang, Li, Chu, Yoo, Bu, de~la Cruz, and Dutta]{miller2019}
M.~Miller, Y.~H. Liang, H.~H. Li, M.~Q. Chu, S.~J. Yoo, W.~Bu, M.~O. de~la Cruz and P.~Dutta, \emph{Physical Review Letters}, 2019, \textbf{122}, 058001\relax
\mciteBstWouldAddEndPuncttrue
\mciteSetBstMidEndSepPunct{\mcitedefaultmidpunct}
{\mcitedefaultendpunct}{\mcitedefaultseppunct}\relax
\EndOfBibitem
\bibitem[Shannon(1976)]{shannon1976}
R.~Shannon, \emph{Acta Crystallographica Section A}, 1976, \textbf{32}, 751--767\relax
\mciteBstWouldAddEndPuncttrue
\mciteSetBstMidEndSepPunct{\mcitedefaultmidpunct}
{\mcitedefaultendpunct}{\mcitedefaultseppunct}\relax
\EndOfBibitem
\end{mcitethebibliography}


\providecommand*{\mcitethebibliography}{\thebibliography}
\csname @ifundefined\endcsname{endmcitethebibliography}
{\let\endmcitethebibliography\endthebibliography}{}
\begin{mcitethebibliography}{6}
\providecommand*{\natexlab}[1]{#1}
\providecommand*{\mciteSetBstSublistMode}[1]{}
\providecommand*{\mciteSetBstMaxWidthForm}[2]{}
\providecommand*{\mciteBstWouldAddEndPuncttrue}
  {\def\EndOfBibitem{\unskip.}}
\providecommand*{\mciteBstWouldAddEndPunctfalse}
  {\let\EndOfBibitem\relax}
\providecommand*{\mciteSetBstMidEndSepPunct}[3]{}
\providecommand*{\mciteSetBstSublistLabelBeginEnd}[3]{}
\providecommand*{\EndOfBibitem}{}
\mciteSetBstSublistMode{f}
\mciteSetBstMaxWidthForm{subitem}
{(\emph{\alph{mcitesubitemcount}})}
\mciteSetBstSublistLabelBeginEnd{\mcitemaxwidthsubitemform\space}
{\relax}{\relax}

\bibitem[Hughes \emph{et~al.}(1981)Hughes, Pailthorpe, and White]{hughes1981}
B.~D. Hughes, B.~A. Pailthorpe and L.~R. White, \emph{Journal of Fluid Mechanics}, 1981, \textbf{110}, 349--372\relax
\mciteBstWouldAddEndPuncttrue
\mciteSetBstMidEndSepPunct{\mcitedefaultmidpunct}
{\mcitedefaultendpunct}{\mcitedefaultseppunct}\relax
\EndOfBibitem
\bibitem[Saffman and Delbruck(1975)]{saffman1975}
P.~G. Saffman and M.~Delbruck, \emph{Proceedings of the National Academy of Sciences of the United States of America}, 1975, \textbf{72}, 3111--3113\relax
\mciteBstWouldAddEndPuncttrue
\mciteSetBstMidEndSepPunct{\mcitedefaultmidpunct}
{\mcitedefaultendpunct}{\mcitedefaultseppunct}\relax
\EndOfBibitem
\bibitem[Saffman(1976)]{saffman1976}
P.~G. Saffman, \emph{Journal of Fluid Mechanics}, 1976, \textbf{73}, 593--602\relax
\mciteBstWouldAddEndPuncttrue
\mciteSetBstMidEndSepPunct{\mcitedefaultmidpunct}
{\mcitedefaultendpunct}{\mcitedefaultseppunct}\relax
\EndOfBibitem
\bibitem[Fischer \emph{et~al.}(2006)Fischer, Dhar, and Heinig]{fischer2006}
T.~M. Fischer, P.~Dhar and P.~Heinig, \emph{Journal of Fluid Mechanics}, 2006, \textbf{558}, 451--475\relax
\mciteBstWouldAddEndPuncttrue
\mciteSetBstMidEndSepPunct{\mcitedefaultmidpunct}
{\mcitedefaultendpunct}{\mcitedefaultseppunct}\relax
\EndOfBibitem
\bibitem[Chisholm and Stebe(2021)]{chisholm2021}
N.~G. Chisholm and K.~J. Stebe, \emph{Journal of Fluid Mechanics}, 2021, \textbf{914}, A29\relax
\mciteBstWouldAddEndPuncttrue
\mciteSetBstMidEndSepPunct{\mcitedefaultmidpunct}
{\mcitedefaultendpunct}{\mcitedefaultseppunct}\relax
\EndOfBibitem
\bibitem[Molaei \emph{et~al.}(2021)Molaei, Chisholm, Deng, Crocker, and Stebe]{molaei2021}
M.~Molaei, N.~G. Chisholm, J.~Y. Deng, J.~C. Crocker and K.~J. Stebe, \emph{Physical Review Letters}, 2021, \textbf{126}, 228003\relax
\mciteBstWouldAddEndPuncttrue
\mciteSetBstMidEndSepPunct{\mcitedefaultmidpunct}
{\mcitedefaultendpunct}{\mcitedefaultseppunct}\relax
\EndOfBibitem
\end{mcitethebibliography}
\bibliographystyle{rsc} 

\end{document}


\title{Supplementary Information for Interfacial Rheology of Lanthanide Binding Peptide Surfactants at the Air-Water Interface}

\author{Stephen A. Crane}
\affiliation{ 
	Department of Chemical and Biomolecular Engineering, University of Pennsylvania, Philadelphia, PA 19104, USA.
}
\author{Felipe Jimenez-Angeles}
\affiliation{ 
	Department of Materials Science and Engineering, Northwestern University, Evanston, IL 60208, USA.
}
\author{Yiming Wang}
\affiliation{ 
	Department of Chemical and Biomolecular Engineering, University of Pennsylvania, Philadelphia, PA 19104, USA.
}
\author{Luis E. Ortuno Macias}
\affiliation{ 
	Department of Chemical Engineering, The City College of New York,  New York, NY 10031, USA.
}
\affiliation{
    Levich Institute, The City College of New York,  New York, NY 10031, USA.
}
\author{Jason G. Marmorstein}
\affiliation{
    Department of Chemistry, University of Pennsylvania,  Philadelphia, PA 19104, USA.
}
\author{Jiayi Deng}
\affiliation{ 
	Department of Chemical and Biomolecular Engineering, University of Pennsylvania, Philadelphia, PA 19104, USA.
}
\author{Mehdi Molaei}
\affiliation{ 
	Department of Chemical and Biomolecular Engineering, University of Pennsylvania, Philadelphia, PA 19104, USA.
}
\author{E. James Petersson}
\affiliation{
    Department of Chemistry, University of Pennsylvania,  Philadelphia, PA 19104, USA.
}
\author{Ravi Radhakrishnan}
\affiliation{ 
	Department of Chemical and Biomolecular Engineering, University of Pennsylvania, Philadelphia, PA 19104, USA.
}
\affiliation{
    Department of Bioengineering, University of Pennsylvania,  Philadelphia, PA 19104, USA.
}
\author{Cesar de la Fuente-Nunez}
\affiliation{ 
	Department of Chemical and Biomolecular Engineering, University of Pennsylvania, Philadelphia, PA 19104, USA.
}
\affiliation{
    Department of Bioengineering, University of Pennsylvania,  Philadelphia, PA 19104, USA.
}
\affiliation{
    Machine Biology Group, Departments of Psychiatry and Microbiology, Institute for Biomedical Informatics, Institute for Translational Medicine and Therapeutics, Perelman School of Medicine, University of Pennsylvania, Philadelphia, PA 19104, USA.
}
\affiliation{
    Penn Insitute for Computational Science, University of Pennsylvania, Philadelphia, PA 19104, USA.
}
\author{ Monica Olvera de la Cruz}
\affiliation{ 
	Department of Materials Science and Engineering, Northwestern University, Evanston, IL 60208, USA.
}
\author{Raymond S. Tu}
\affiliation{ 
	Department of Chemical Engineering, The City College of New York,  New York, NY 10031, USA.
}
\author{Charles Maldarelli}
\affiliation{ 
	Department of Chemical Engineering, The City College of New York,  New York, NY 10031, USA.
}
\affiliation{
    Levich Institute, The City College of New York,  New York, NY 10031, USA.
}
\author{Ivan J. Dmochowski}
\affiliation{
    Department of Chemistry, University of Pennsylvania,  Philadelphia, PA 19104, USA.
}
\author{Kathleen J. Stebe}
\email{kstebe@seas.upenn.edu}
\affiliation{ 
	Department of Chemical and Biomolecular Engineering, University of Pennsylvania, Philadelphia, PA 19104, USA.
}

\date{April 28, 2024}

\maketitle

 \section{Resolution of Mean Squared Displacement}
\begin{figure}[H]
 \centering
 \includegraphics[height=5cm]{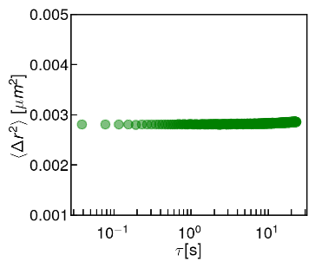}
 \caption{The MSD resolution of the microscope used in this work is based on the perceived motion of static particles is about 0.003 $\mu m^2$.}
 \label{sfgr1}
\end{figure}

To find the minimum MSD our microscope can measure, dried polystyrene particles ($2a=1\mu m$) were placed directly on glass coverslip. The apparent motion of these fixed particles was tracked. The resulting MSD of the particles is shown in Fig. \ref{sfgr1}. This nearly flat MSD vs lag-time shows the minimum MSD we can resolve.

\section{Relationships between diffusivity and Bq}
\begin{figure}[H]
 \centering
 \includegraphics[height=5cm]{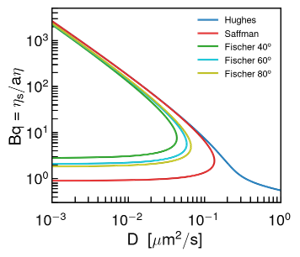}
 \caption{Various expressions relating D to Bq all give similar results. }
 \label{sfgr2}
\end{figure}

Three relationships between particle diffusivity, D, and surface viscosity, made non-dimensional in Bq, were considered. In each case the diffusivity is related to a drag coefficient through the Stokes-Einstein relation:
\begin{equation}
  D = \frac{k_BT}{4\pi\eta a\Lambda}
  \label{seq1}
\end{equation}
where $\eta$ is the viscosity of the subphase, $a$ is the particle radius, and $\Lambda$ is the drag coefficient. The following expressions for $\Lambda$ are taken from Hughes, et. al.\cite{hughes1981}, Saffman and Delbruck\cite{saffman1975,saffman1976}, and Fischer et. al.\cite{fischer2006}

\begin{equation}
  \Lambda_{Hughes}=\left[\frac{1}{Bq}\left(\ln{2Bq}-\gamma+\frac{4}{\pi Bq}-\frac{1}{2Bq^2}\ln{2Bq} \right)\right]
  \label{seq2}
\end{equation}

\begin{equation}
  \Lambda_{Saffman}=\left[\frac{1}{Bq}\left(\ln{2Bq}-\gamma\right)\right]
  \label{seq3}
\end{equation}

\begin{equation}
  \Lambda_{Fischer}=\left[\frac{1}{Bq}\left(\ln{Bq\sin{\theta}}-\gamma\right)\right]
  \label{seq4}
\end{equation}
where $\gamma$ is the Euler-Mascheroni constant and $\theta$ is the contact angle of the particle with the fluid interface. Fig. \ref{sfgr2} shows Bq as calculated from D using these different expressions. For this work we chose to use the Hughes relationship since it covers the largest range of Bq.

\section{Interfacial Rheology of Control Interfaces}
\begin{figure}[H]
 \centering
 \includegraphics[height=5cm]{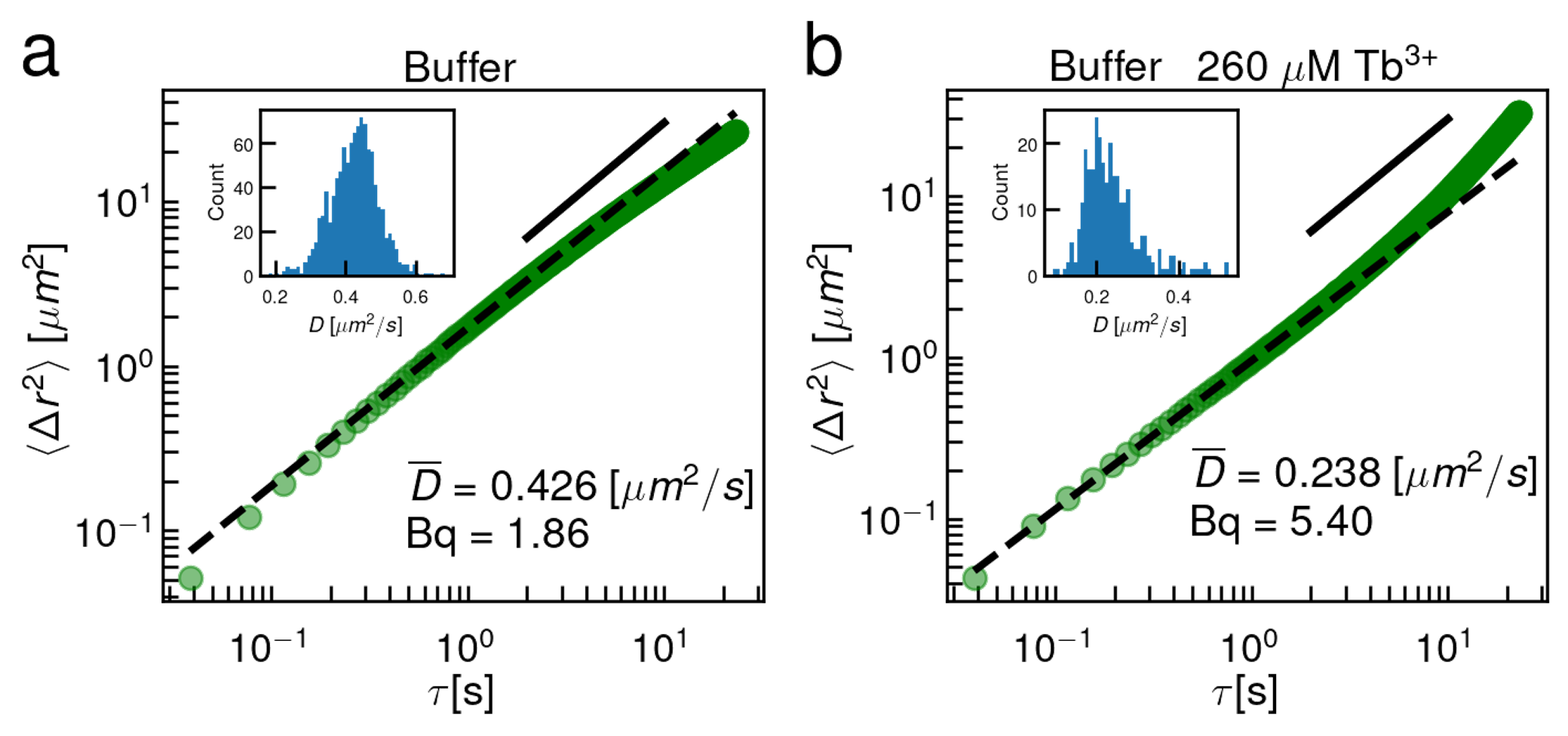}
 \caption{Addition of Tb$^{3+}$ to a buffer solution increases surface viscosity slightly. EMSD of probe particles ($2a=1\mu m$) at the air-water interface in the presence of bulk solution containing (a) 50 mM MES and 100 mM NaCl buffer solution; (b) buffer solution and 260 $\mu$M Tb$^{3+}$. Insets in (a) and (b) show the distribution of particle diffusivities. The black dashed lines are fits of the form MSD=4D$\tau$. The solid black lines are guides to show a power law with an exponent of 1. }
 \label{sfgr3}
\end{figure}

To ensure the results we measure with PEPS and Tb$^{3+}$ present are due to the adsorbed layer of PEPS, two controls are used. We measured the surface viscosity of buffer solution that contained no PEPS and a buffer solution that contained 260 $\mu$M Tb$^{3+}$ (Fig. \ref{sfgr3}). We find that the surface viscosity increases slightly when Tb$^{3+}$ is present at 260 $\mu$M, however, the measured Bq = 5.40 is significantly smaller than the Bq measured for solutions of PEPS and Tb$^{3+}$ at high concentrations. Therefore, we are confident that the Bq we measure for PEPS containing solutions is largely attributable to the interfacial PEPS layer and not impurities in our buffer or Tb$^{3+}$ solution.

\section{Correlated Displacement Velocimetry Measurement Resolution}
\begin{figure}[H]
 \centering
 \includegraphics[height=16cm]{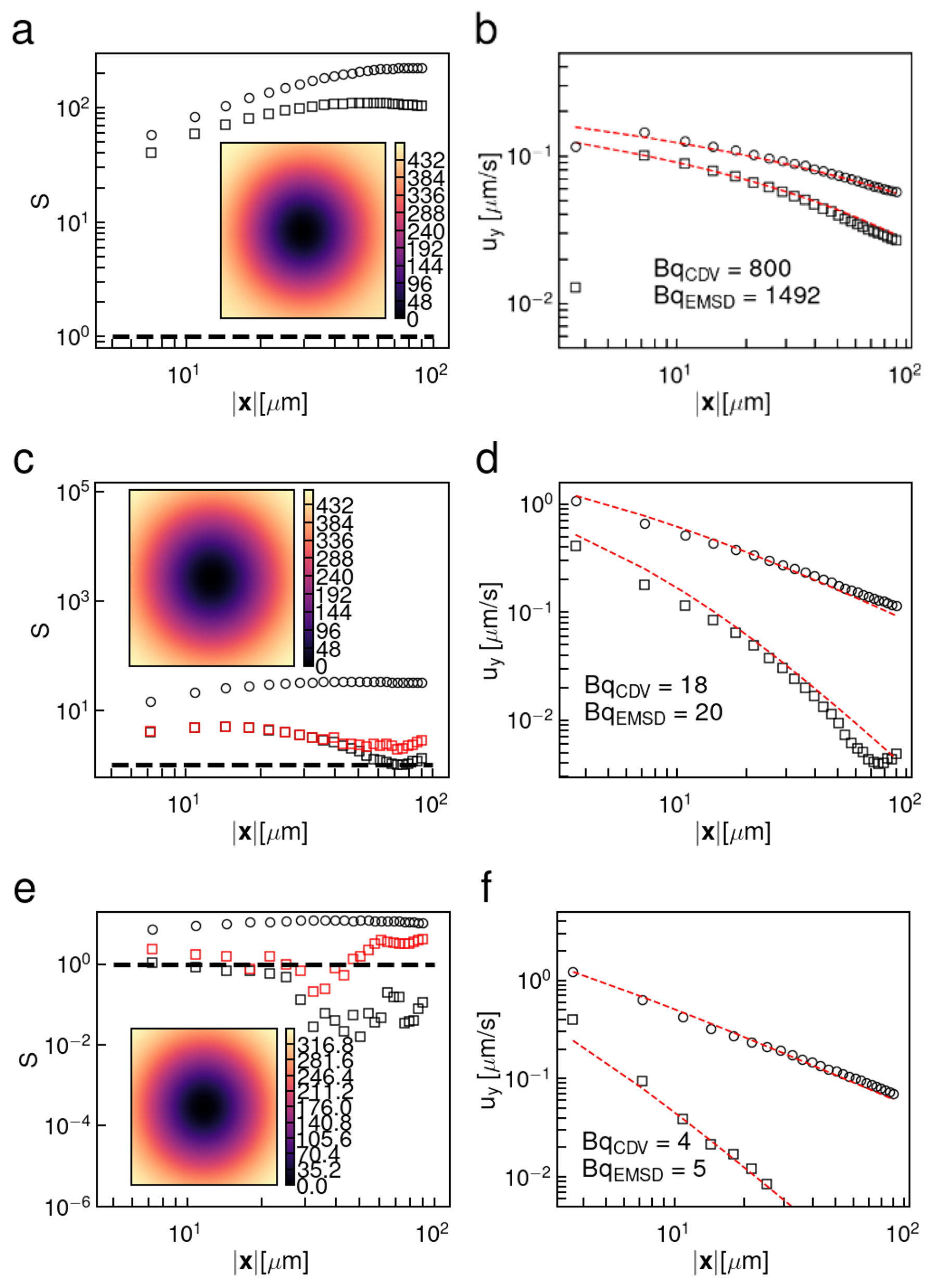}
 \caption{Resolution analysis for representative Bq interfaces. (a, c, e) Significance metric, $S$, for decreasing Bq. Inset is the number of datapoints at each location on the interface in 1000s of datapoints. The circles represent the y-directed displacements along the y axis, the black squares represent the y-directed displacements along the x axis. The red squares represent the total displacement. (b, d, f) Fits to the velocity decays for increasing Bq demonstrating how the data is cutoff when it is below resolution.}
 \label{sfgr4}
\end{figure}

To ensure that the displacements calculated by correlated displacement velocimetry (CDV) are resolved above noise from Brownian motion the following inequality must be true:
\begin{equation}
  U(|\mathbf{x}|)\gg\frac{\langle\Delta r^2\rangle^{1/2}}{\sqrt{N}}
  \label{seq5}
\end{equation}
where $U$ is the displacement $(U=u_y\tau)$,$\langle\Delta r^2\rangle^{1/2}$ is the mean squared displacement at the lag time that the CDV is constructed, and N is the number of data points in a bin at the location $|\mathbf{x}|$. In words, the average Brownian noise is given by the MSD of the particles. We divide this average noise by the number of data points and take the square root to calculate a value akin to the standard error of the mean Brownian motion. If the displacement is larger than this value, then it can be resolved above the noise level. We define a significance criterion, $S$:

\begin{equation}
    S\equiv\frac{U(|\mathbf{x}|)\sqrt{N}}{\langle\Delta r^2\rangle^{1/2}}\gg1
  \label{seq6}
\end{equation}
and plot the left side of this expression versus $|\mathbf{x}|$ to determine whether sufficient data have been binned for displacements at given locations in the interface to be considered significant. Figure S\ref{sfgr4} shows examples of these significance plots.

Our data shows for large Bq the resolution criterion is always met. This is because the Brownian noise is considerably smaller for such viscous interfaces. As Bq decreases, the Brownian noise increases and the velocity decays faster. As a result, the resolution criterion begins to fail. Still, the flow fields are largely well-resolved. The red squares, which represent the total displacement (Fig. \ref{sfgr4}c, e), typically meet the criterion despite the y-directed displacements not being strong enough. This implies that while the entire flow field may not be quantitatively accurate for fitting to find Bq, the streamlines and velocity magnitudes are accurate. Lastly, we highlight that the y-directed velocities along the y axis are always above the resolution criterion which allows Bq fitting to always take place. Generally, the y-directed velocities along the x axis are most sensitive to Bq, but when these velocities are not well resolved, the y axis decays are sufficient.

\section{CDV flow field corrections}
\begin{figure}[H]
 \centering
 \includegraphics[height=10cm]{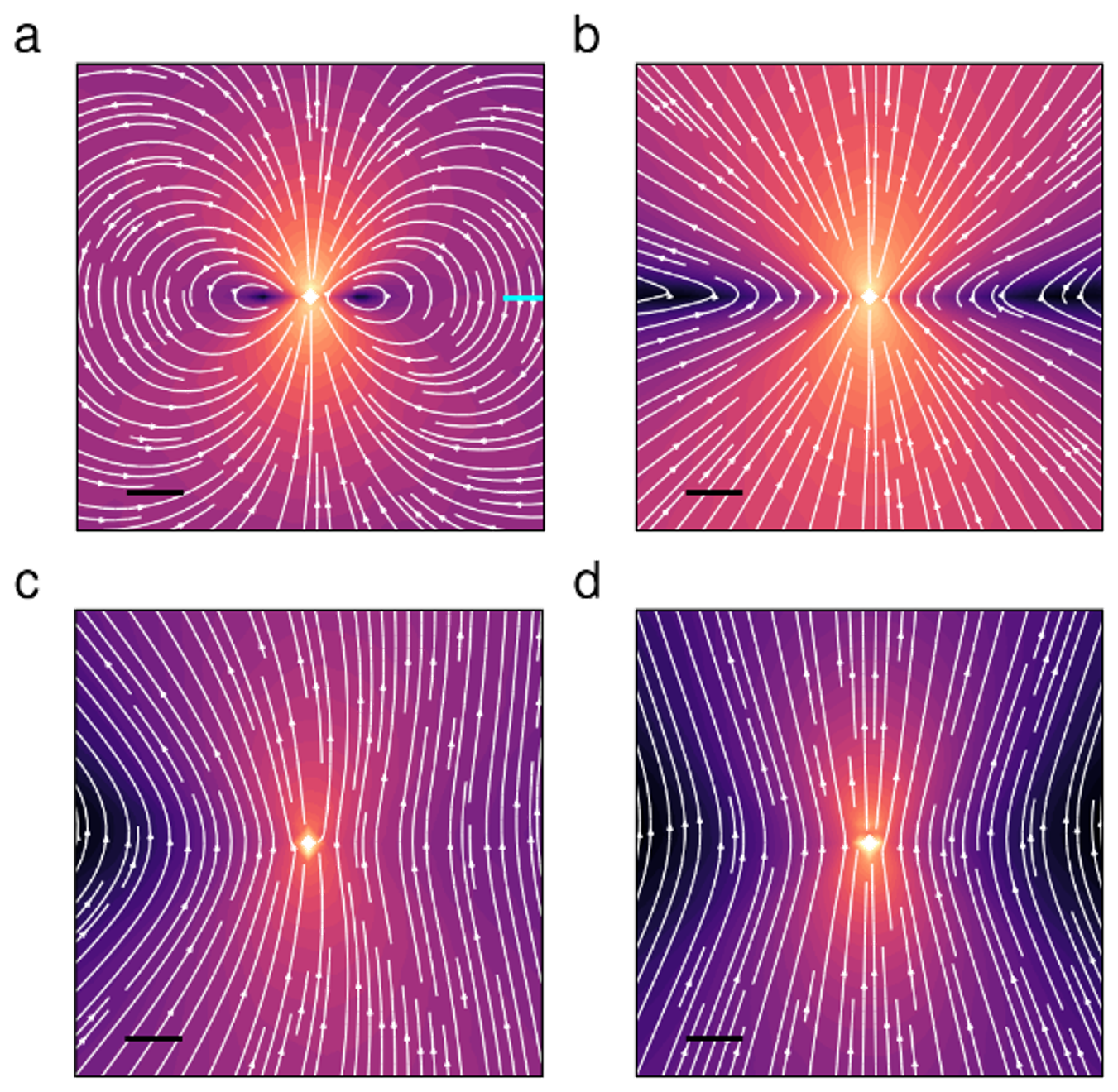}
 \caption{Flow field (a) before drift correction and (b) after drift correction. The average y-directed velocity is taken from the points highlighted by the cyan bar and the multiplier for this mean velocity is 1.15. Flow field (c) before symmetrization and (d) after symmetrization. Scale bars are 20 $\mu$m.}
 \label{sfgr5}
\end{figure}

In most cases, drift affects the data which causes the flow field to show a recirculatory flow. Since such flow is not expected, as shown in Fig. \ref{sfgr5}, we apply a correction to the flow field by subtracting the average y-directed velocity along the x axis far from the origin multiplied by a constant close to 1.0. The exact value of the constant is found by looking at the resulting velocity decays and picking a value where those decays are smoothest. In addition, there are sometimes significant asymmetries which we remove by averaging. This averaging is done by taking the average of the velocities at the same location relative to the x and y axes. We perform this correction on all flow fields regardless of whether they are asymmetric or not. Some examples of these corrections are shown in Fig. \ref{sfgr5}.

\section{Correlated Displacement Velocimetry}
\begin{figure}[H]
 \centering
 \includegraphics[height=10cm]{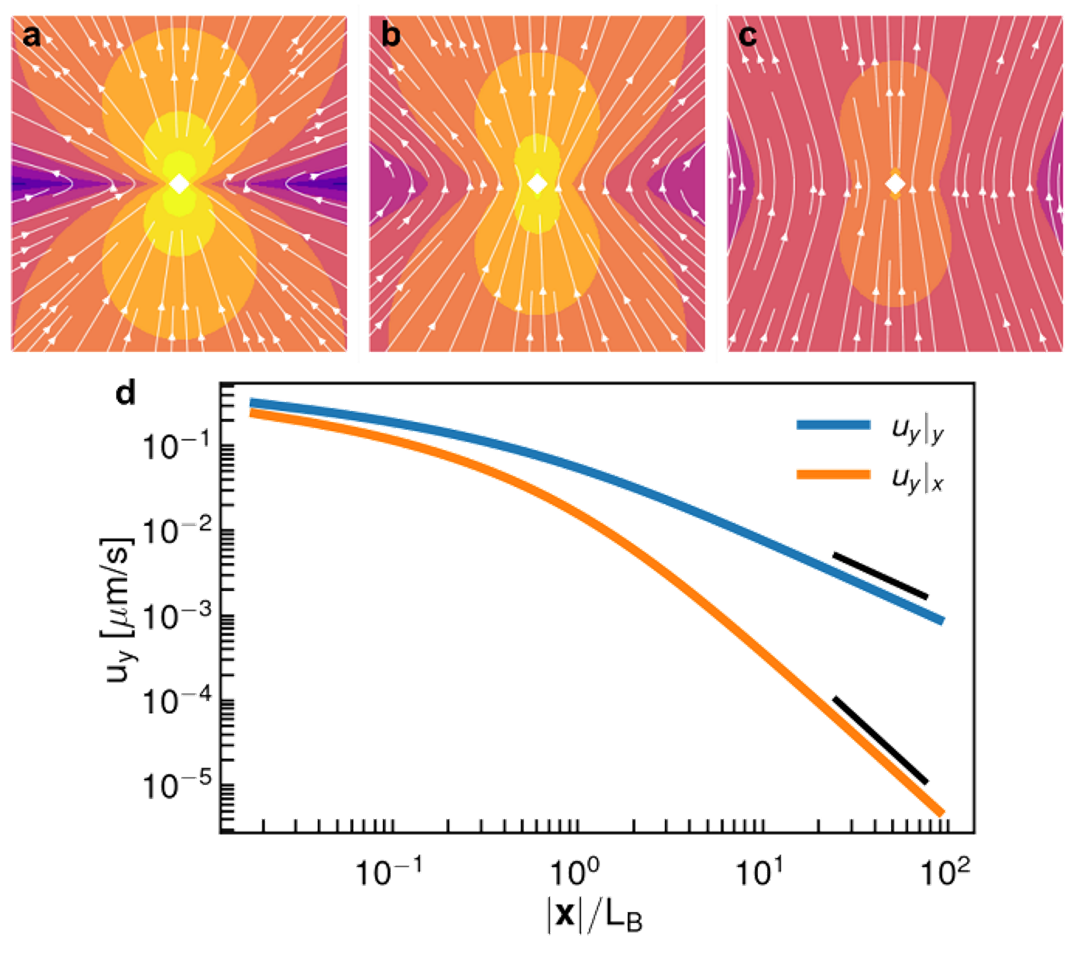}
 \caption{Theoretical flow fields for increasing Bq show a widening waist; y-directed velocity along the x and y axes decays as $|\mathbf{x}|^{-2}$ and $|\mathbf{x}|^{-1}$, respectively, for separation distances larger than $L_B$. Flow fields with (a) Bq = 3; (b) Bq = 30; (c) Bq = 300. The black lines are guides to show $|\mathbf{x}|^{-2}$ and $|\mathbf{x}|^{-1}$ decays. (d) y-directed velocity along the x axis (orange) and y axis (blue) versus separation distance divided by the Boussinesq length $L_B$.}
 \label{sfgr6}
\end{figure}

The theoretical flow fields that were derived for a point force in the interface\cite{chisholm2021} are shown in the top row of Fig. \ref{sfgr6}. Bq increases from 3 (Fig. \ref{sfgr6}a) to 30 (Fig. \ref{sfgr6}b) to 300 (Fig. \ref{sfgr6}c). The streamlines show a waist which widens with increasing Bq. The colormap which depicts the velocity magnitude shows that there is slower decay for larger Bq. The decay of the y-directed velocities along the x and y axes as a function of the separation distance $|\mathbf{x}|$ divided by the Boussinesq length $(L_B=\eta_s/\eta)$ is depicted in Fig. \ref{sfgr6}d. Once $|\mathbf{x}|/L_B$ becomes greater than $\sim$5, the decay of the y-directed velocity along the y-axis becomes proportional to $|\mathbf{x}|^{-1}$ and the decay of the y-directed velocity along the x-axis is proportional to $|\mathbf{x}|^{-2}$. We extract y-directed velocities from measured flow fields\cite{molaei2021} and fit their decays to these theoretical forms to find the Bq of the interface. The fitting is achieved by first finding the magnitude of the point force which shifts the entire decay curve up or down. Next, LB¬ is found using interval halving around an initial guess based on the single point value for Bq until a minimum sum of square error is found. $L_B$ is multiplied by 2.0 to get Bq since Bq = $L_B/a$ and a = 0.5 $\mu$m.

\section{Simulation Snapshots}
The simulation box, along with some snapshots are shown below. Note in Fig. \ref{sfgr7}b and d the difference in the location of most excess Tb$^{3+}$. For LBT$^{5-}$ (Fig. \ref{sfgr7}b) the excess Tb$^{3+}$ localize in the interfacial layer, whereas for LBT$^{3-}$ (Fig. \ref{sfgr7}d) the excess Tb$^{3+}$ are more distributed throughout the solution.

\begin{table}[H]
    \centering
    \caption{MD Simulation Parameters. The first column is the system label, the second column is the peptide type, the third column is the Tb$^{3+}$ to peptide concentration ratio, and the fourth column is the peptide surface area density in $molecules/\AA^2$. $N_P$, $N_w$, $N_{Tb}$, $N_{Na}$, and $N_{Cl}$ represent the number of peptide molecules, water molecules, Tb$^{3+}$ ions, Na$^{+}$ ions, and Cl$^{-}$ ions, respectively.}
    
    \begin{tabular} {|| c | c | c | c | c | c | c | c | c ||} \hline
    System & Peptide & $\frac{[Tb^{3+}]}{[LBT]}$ & $\Gamma _P (molecules/\AA^2)$ & $N_P$ & $N_w$ & $N_{Tb}$ & 
    $N_{Na}$ & $N_{Cl}$  \\ \hline
    1 & $LBT^{5-}$ & 2 & 0.0054 & 108 & 41980 & 216 & 96 & 204\\ \hline
    2 & $LBT^{5-}$ & 2 & 0.0040 & 80 & 44550 & 160 & 96 & 176\\ \hline 
    3 & $LBT^{5-}$ & 1.75 & 0.0040 & 80 & 44550 & 140 & 96 & 116\\ \hline 
    4 & $LBT^{5-}$ & 1.50 & 0.0040 & 80 & 44550 & 120 & 136 & 96\\ \hline 
    5 & $LBT^{5-}$ & 1.25 & 0.0040 & 80 & 44550 & 100 & 196 & 96\\ \hline 
    6 & $LBT^{5-}$ & 1 & 0.0040 & 80 & 44550 & 80 & 256 & 96\\ \hline 
    7 & $LBT^{3-}$ & 2 & 0.0051 & 102 & 42436 & 204 & 96 & 402\\ \hline 
    8 & $LBT^{3-}$ & 1 & 0.0051 & 102 & 42436 & 102 & 300 & 300\\ \hline 
    \end{tabular}
    \label{stabl1}
\end{table}

\begin{figure}[H]
 \centering
 \includegraphics[height=16cm]{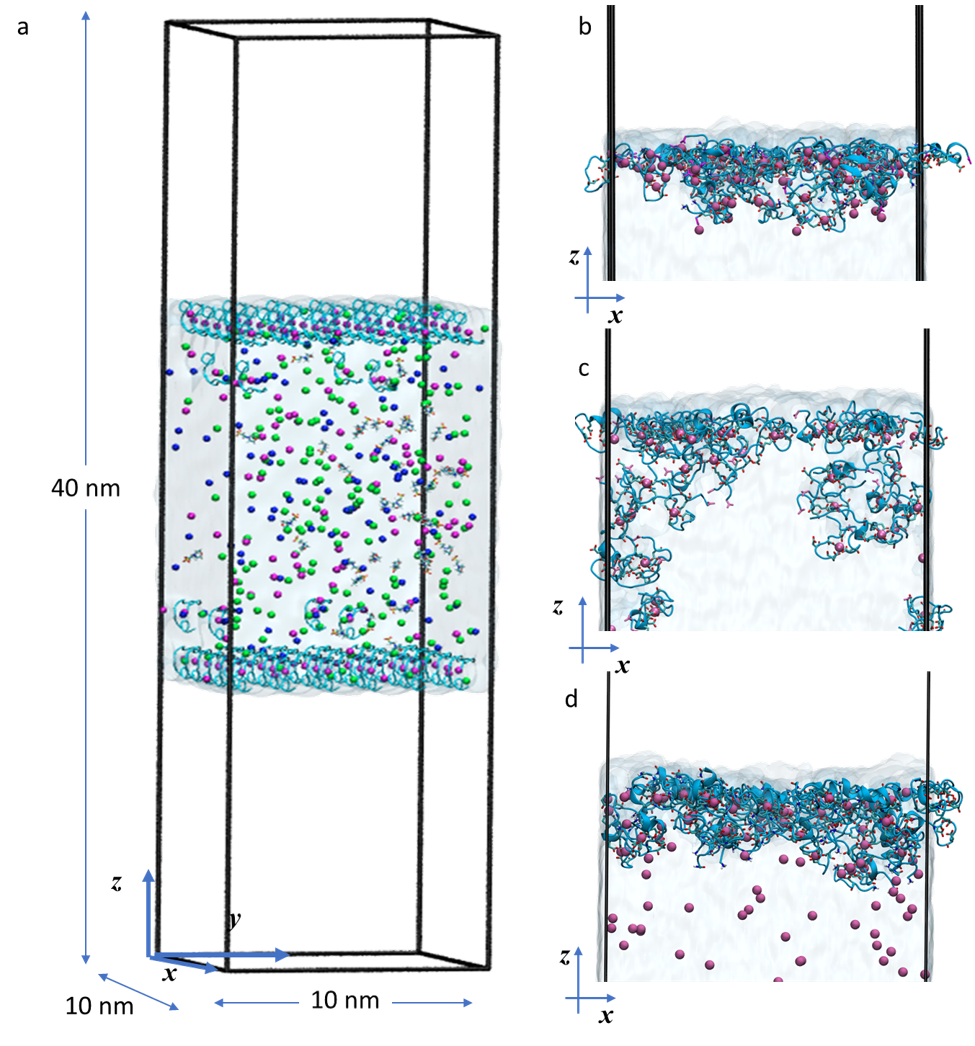}
 \caption{(a) The simulation box with PEPS in their starting position. Tb$^{3+}$ ions are in purple, Cl$^{-}$ ions in green, and Na$^{+}$ ions in blue. Side views of the interface for (b) Tb$^{3+}$:LBT$^{5-}$ 2:1; (c) Tb$^{3+}$:LBT$^{5-}$ 1:1; (d) Tb$^{3+}$:LBT$^{3-}$ 2:1. }
 \label{sfgr7}
\end{figure}

\section{Viscous Rheology of LBT$^{3-}$}
\begin{figure}[H]
 \centering
 \includegraphics[height=16cm]{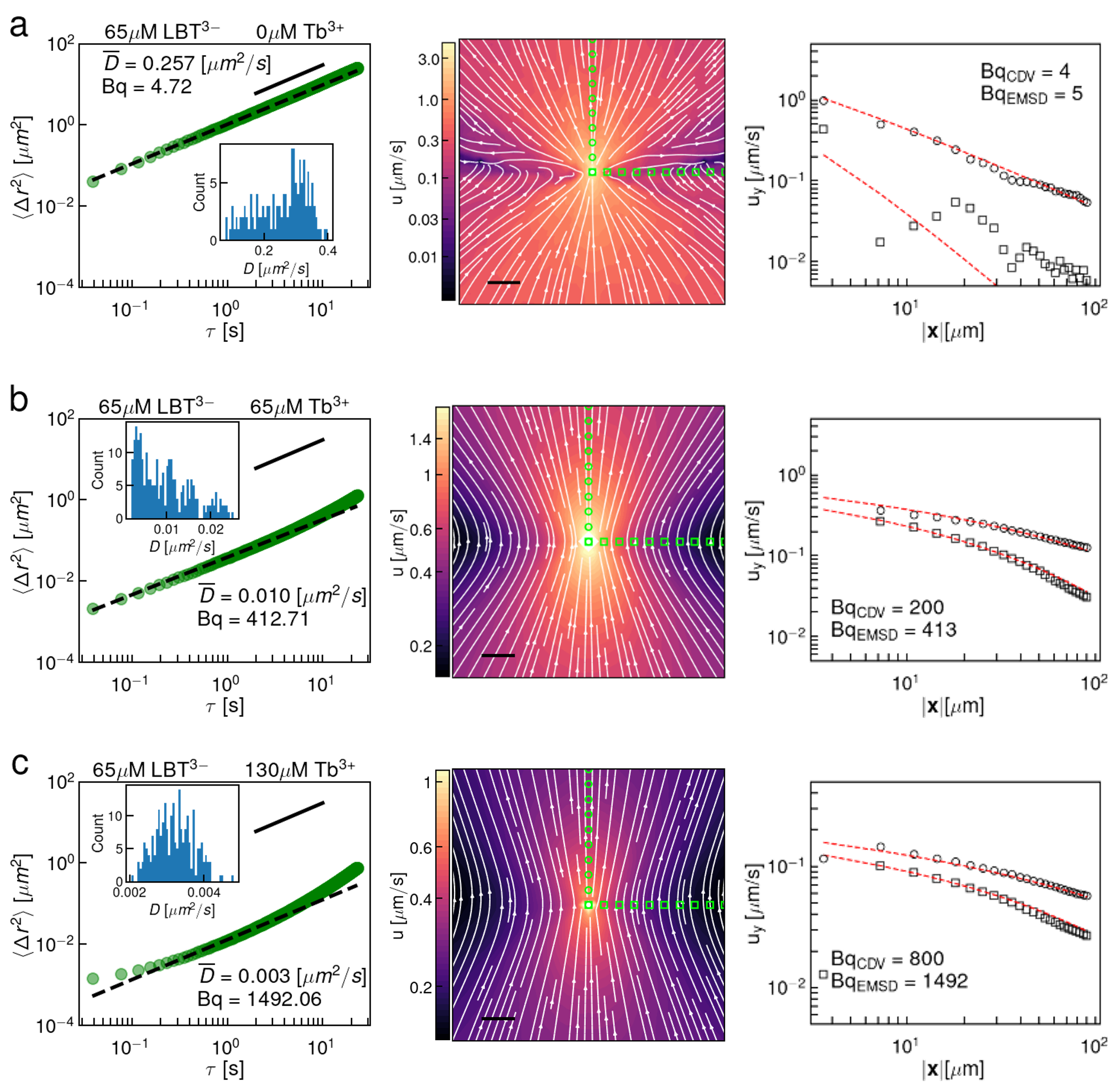}
 \caption{CDV for LBT$^{3-}$. (Left column) EMSD of probe particles $(2a=1\mu m)$ at the air-water interface. Insets show the distribution of particle diffusivities. The black dashed lines are fits of the form MSD=4D$\tau$. The solid black lines are guides to show a power law with an exponent of 1. (Middle column) flow fields constructed from CDV. Scale bars are 20 $\mu$m. (Right column) y-directed velocity decays along the y-axis (circles) and x-axis (squares) fits from the flow field functional form, used to find Bq, are shown in red dashed lines.}
 \label{sfgr8}
\end{figure}

\section{Aggregation at the air water interface}

\begin{figure}[H]
 \centering
 \includegraphics[height=5cm]{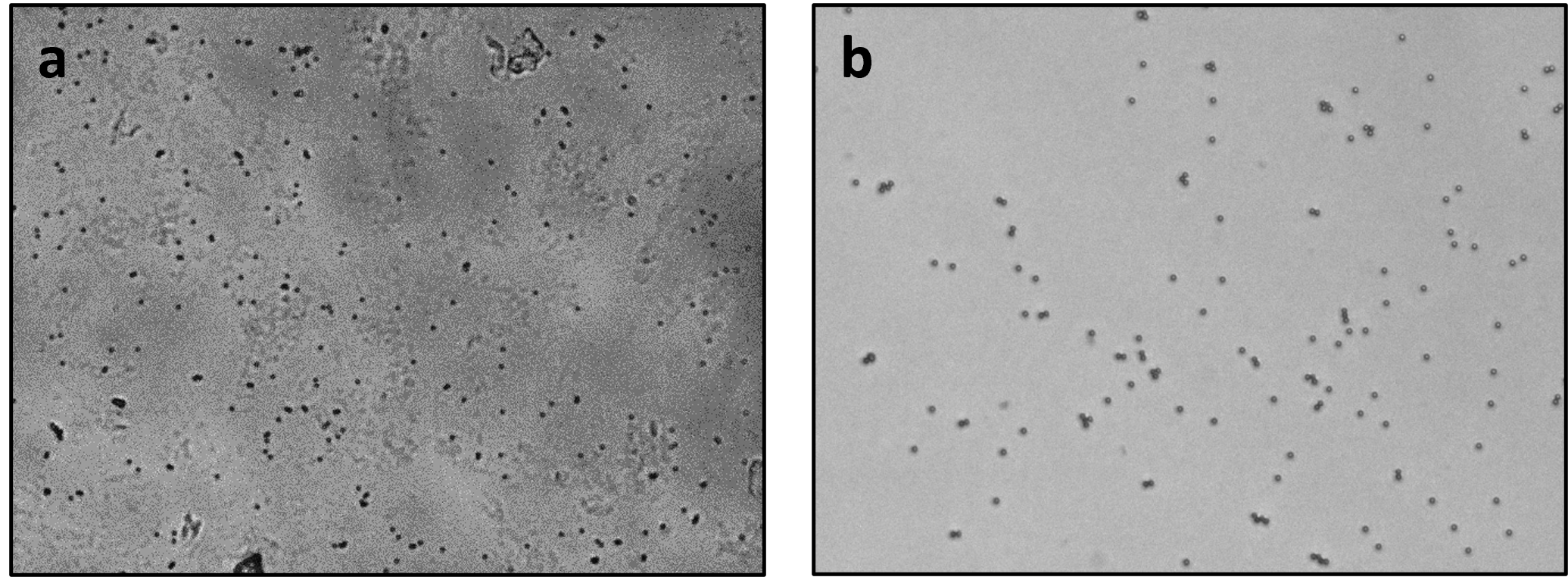}
 \caption{Aggregates present in the air-water interface upon addition of Al$^{3+}$. (Left) Micrograph of the air-water interface with 400 $\mu$M Al$^{3+}$ and 100 $\mu$M LBT$^{3-}$ containing solution. (Right) Micrograph of typical interface of solution containing Tb$^{3+}$ and PEPS over the range of concentrations studied. The black dots are 2a = 1 $\mu$m polystyrene beads that are trapped in the interface which can be used to determine the scale.}
 \label{sfgr9}
\end{figure}

\section{Calculating Basin Free Energy}
The free energy of each basin can be calculated according to:
\begin{equation}
    \langle F_{basin} \rangle = -k_BT\log \left ( \int_{s_1(b)}^{s_1(a)}\int_{s_2(d)}^{s_2(c)}e^{-F(s_1,s_2)/k_BT}ds_1ds_2 \right )
  \label{seq7}
\end{equation}
where $\langle F_{basin} \rangle$ is the free energy value integrated over a given two-dimensional area covered by collective variables $s_1$ and $s_2$ which corresponds to the coordination number (CN) of Tb$^{3+}$ and the root mean square deviation of $\alpha$ carbon $(C_\alpha$-RMSD) of the binding loop domain of LBT peptide respectively. $k_B$ is the Boltzmann constant, $T$ is the temperature. 
\begin{figure}[H]
 \centering
 \includegraphics[height=7cm]{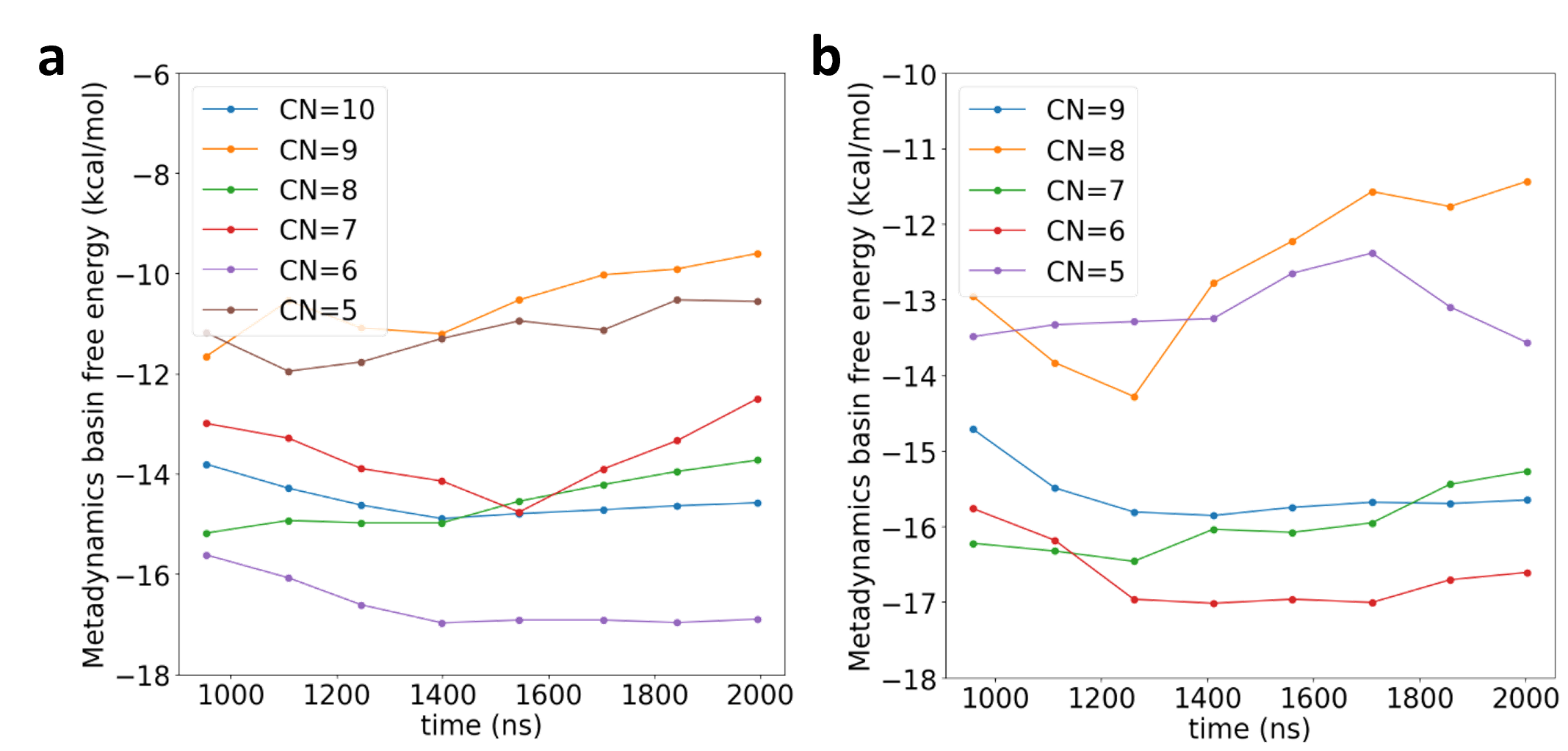}
 \caption{Plot of free energies for different basins (CN=5-10) of (a) LBT$^{5-}$:Tb$^{3+}$ and (b) LBT$^{3-}$:Tb$^{3+}$ binding structures as a function of simulation time, for checking the convergence of the enhanced sampling.}
 \label{sfgr10}
\end{figure}

\section{Metadynamics Violin Plots}
\begin{figure}[H]
 \centering
 \includegraphics[height=6cm]{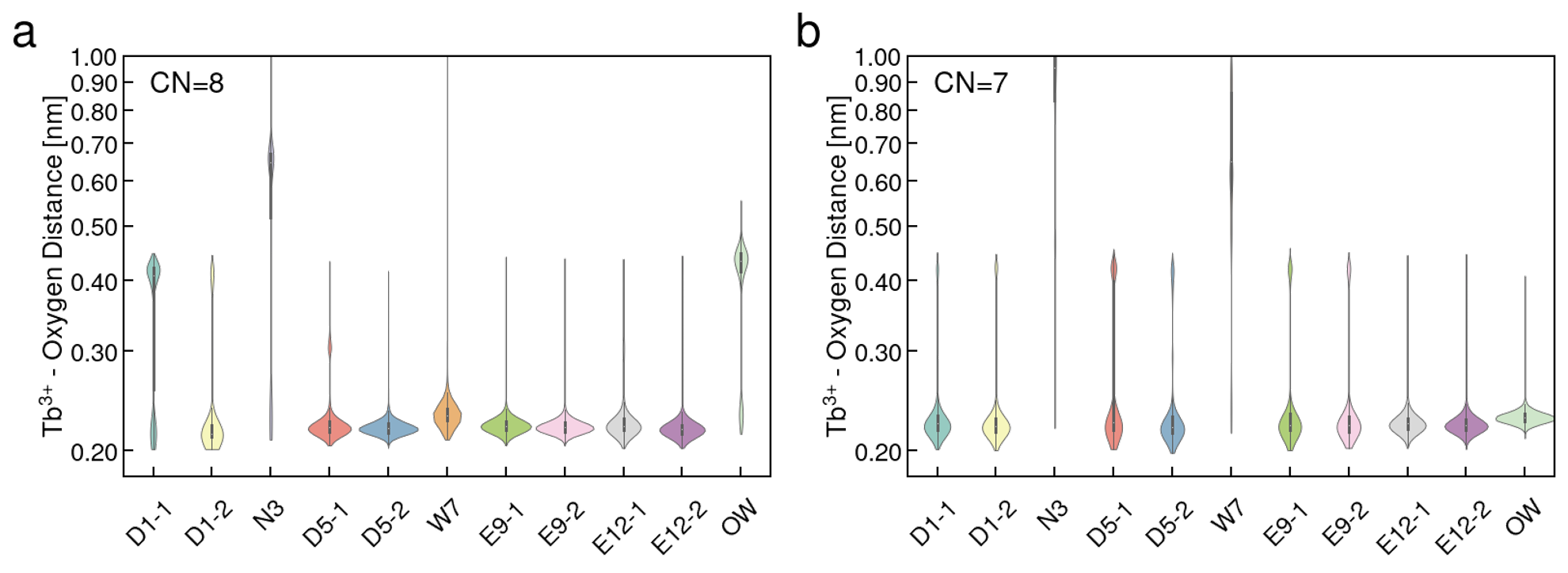}
 \caption{Violin plots for (a) CN=8, LBT$^{5-}$:Tb$^{3+}$ and (b) CN=7 LBT$^{3-}$:Tb$^{3+}$.}
 \label{sfgr11}
\end{figure}

\section{Aggregation Videos}
Supplemental videos can be found as separate files.

\bibliography{refs} 
\bibliographystyle{rsc}